\documentclass[%
 reprint,
 superscriptaddress,
 amsmath,amssymb,
 aps,
]{revtex4-2}

\usepackage{graphicx}% Include figure files
\usepackage{orcidlink}
\usepackage{hyperref}% add hypertext capabilities
\hypersetup{
    colorlinks=true,
    linkcolor=red,
    citecolor=green,
    anchorcolor=black,
    urlcolor=magenta
}
\usepackage{siunitx}% for SI units
\usepackage{xcolor}% for colours
\usepackage{comment}
\usepackage{slashed}

\renewcommand{\d}{\mathrm{d}}

\begin{document}

\preprint{APS/123-QED}

\title{Scalar-Mediated Inelastic Dark Matter as a Solution to Small-Scale Structure Anomalies}
\author{Zihan~Wang}
\email{zihan.wang@queens.ox.ac.uk}
\affiliation{Department of Physics, University of Oxford, Keble Road, Oxford, OX1 3PU, UK}

\begin{abstract}
    We propose a novel Self-Interacting Dark Matter (SIDM) model mediated by a light leptophilic scalar boson to solve long-standing small-scale structure problems within the $\Lambda$CDM framework. Small-scale anomalies such as the core--cusp problem challenge the traditional CDM framework. Therefore, we introduce a scalar-mediated SIDM model as an alternative to CDM that naturally addresses these problems. It also exhibits $p$-wave-suppressed annihilation and avoids constraints from the Cosmic Microwave Background (CMB). In our model, we assume pseudo-Dirac dark matter with a small mass splitting of order $10^2$ eV to ensure kinematic scattering suppression in satellites. We also introduce a dimension-5 transition magnetic dipole operator to satisfy Big Bang Nucleosynthesis (BBN) requirements. It allows the decay $\chi_2 \rightarrow \chi_1 \gamma$, so the excited-state abundance is primarily depleted by exothermic self-interactions ($\chi_2 \chi_2 \to \chi_1 \chi_1$) in the early universe. This ensures that the excited-state population is negligible for structure formation at the dipole scale $\Lambda_{\text{eff}} \gtrsim 10^7$ GeV. The model is made inelastic by a dark discrete $\mathbb{Z}_2$ symmetry that prevents tree-level elastic scattering. Treating the scattering dynamics non-perturbatively in a Schr\"odinger framework yields a relatively small residual resonant benchmark window ($m_\chi \approx 40$ GeV, $\Delta m \approx 100$ eV, $m_\phi \approx 20$ MeV). In such a system, the kinematic threshold minimizes interactions in ultra-faint Milky Way satellites, while resonant effects produce large cross-sections, $\sigma/m \sim \mathcal O(10)$ cm$^2$/g, in field dwarf galaxies. Current direct-detection prospects are weak, and the leptophilic scalar mediator keeps the model away from existing nuclear recoil (NR) constraints. However, the low-threshold dipole operator provides a potential discovery channel for future xenon-based experiments.
\end{abstract}

\maketitle

\section{Introduction}
\label{sec:intro}

The classical cosmological model Lambda Cold Dark Matter ($\Lambda$CDM) has been successful in explaining the large-scale structure of the universe~\cite{Planck2018}. Its predictions agree well with observations ranging from temperature anisotropies in the Cosmic Microwave Background (CMB) to galaxy cluster distributions. However, on sub-galactic scales, with distances $\lesssim 10$ kpc and halo masses $\lesssim 10^{11} M_\odot$, significant deviations from collisionless $N$-body simulations and astrophysical observations appear~\cite{Bullock2017}. Classified as ``small-scale crises,'' these deviations suggest that the dark sector may depart from the collisionless assumption~\cite{Springel:2005nw}. The anomalies include: (i) the core--cusp problem: CDM simulations predict dark matter halos with high central-density cusps ($\rho \propto r^{-1}$)~\cite{Navarro1997}, while rotation curves in dwarf and Low Surface Brightness (LSB) galaxies favor low-density cores ($\rho \propto r^0$)~\cite{Oh2011, Moore1994, deBlok2010}; (ii) the diversity problem: galaxies with similar maximum circular velocities ($V_\text{max}$), corresponding to similar halo masses, show large scatter in their inner rotation curves ($V_\text{2kpc}$)~\cite{Oman2015}; (iii) the too-big-to-fail problem: simulations predict a large population of massive subhalos in Milky Way-like systems that is not observed~\cite{Klypin1999, Moore1999}. Some subhalos may be dark due to baryonic feedback, and the dynamical properties of the most massive subhalos differ from the observed kinematics of the brightest satellite galaxies~\cite{BoylanKolchin2011, GarrisonKimmel2014}.

Self-Interacting Dark Matter (SIDM)~\cite{Spergel2000, Tulin2018} is one promising approach to these issues. If dark matter particles scatter with a large cross-section ($\sigma/m \sim 1$ cm$^2$/g), heat transfer from the dynamically hot outer halo to the colder interior can thermalize the central region, reduce central density, and transform cusps into cores~\cite{Rocha2013, Peter2013}. SIDM interaction rates and halo concentration dynamics can also explain the diversity of rotation curves~\cite{Kamada2017, Ren2019}. However, a robust SIDM model must satisfy velocity-dependent constraints. Observations of galaxy clusters, including halo shapes and mergers such as the Bullet Cluster, limit the high-velocity cross-section ($v \sim 10^3$ km/s) to $\sigma/m \lesssim 0.6$ cm$^2$/g~\cite{Markevitch2004, Randall2008, Harvey2015, ODonnell:2025pkw}. For dwarf galaxies ($\sim 30\text{--}60$ km/s), values around $\sigma/m \sim 1\text{--}10$ cm$^2$/g are needed to significantly modify density profiles~\cite{Zavala2013}. Ultra-faint Milky Way satellites (e.g., Draco, 10--20 km/s) provide some of the strongest upper limits on self-interaction~\cite{Valli2018}. Excessive scattering in these systems may trigger the Nishikawa instability, runaway gravothermal collapse, and central densities much larger than those initially studied in~\cite{Nishikawa2020}. Taken together, observations require an effective model that is active at intermediate velocities (e.g., dwarfs) but suppressed at high (clusters) and very low (satellites) velocities. Light mediators with a Yukawa potential can provide cluster-scale suppression while exhibiting Rutherford-like behavior, $\sigma \propto v^{-4}$~\cite{Tulin2013}.

However, as we can see with only a simple elastic cross-section, there are problems when it comes to getting the satellite and dwarf requirements to match. Inelastic scattering is therefore an interesting solution. It sets a threshold of kinematics: The endothermic channel ($\chi_1 \chi_1 \rightarrow \chi_2 \chi_2$) is kinematically forbidden by the mass splitting $\Delta m$ below the endothermic velocity $v_\text{th} \sim \sqrt{8\Delta m/m_\chi}$. This threshold acts as a switch that turns off interactions in the coldest halos~\cite{TuckerSmith2001, Slatyer2009}. The idea mainly applies to vector mediator known as dark photons~\cite{ArkaniHamed2009}. However, vector mediators create unavoidable tension with cosmological variables. The thermal annihilation of vector-coupled dark matter are $s$-wave, as energy is injected into the plasma in the recombination epoch and warps the CMB~\cite{Planck2018, Slatyer:2015jla,Lopez-Honorez:2013cua}. To escape these constraints, we usually need to fine-tune the mass splittings or construct asymmetric dark matter models~\cite{Kaplan2009, Petraki2013}. Therefore,we present a scalar-mediated inelastic dark matter model in this work. The answer is indeed enough to satisfy CMB constraints since we can plug the vector mediator into a scalar, which is already a $p$-wave suppressed annihilation without any fine-tuning. To prevent tree-level elastic scattering, we apply a simple $\mathbb{Z}_2$ symmetry in the dark sector. Moreover, to prevent violating any rules of Big Bang Nucleosynthesis (BBN), we present a leptophilic coupling structure to a scalar mediator. 

The paper is organized as follows. The theoretical framework and dipole mechanism are discussed in Sec.~\ref{sec:theory}. Then, we discuss the non-perturbative scattering phenomenology in Sec.~\ref{sec:scattering}. Sec.~\ref{sec:cosmo} describes cosmological consistency. Sec.~\ref{sec:results} provides numerical results. Also, we constrained our model with direct search experiments in Sec.~\ref{sec:dd_detailed}, and pointed out future detection possibilities. Finally, we discuss and conclude this paper in Sec.~\ref{sec:conclusion}. In addition, for detailed derivations of the coupled-channel scattering formalism, please refer to Appendix~\ref{app:scattering}. 

\section{Theoretical Framework}
\label{sec:theory}

We predict a dark sector that is outside the Standard Model (SM), interacting with SM weakly via necessary portal effects. The particles in this sector consist of Dirac fermion $\chi$ and light real scalar boson $\phi$, giving a kinetic term:
\begin{equation}
    \mathcal{L}_{\text{kinetic}}=\bar\chi(\mathrm i\slashed\partial-m_\chi)\chi-\frac{\Delta m}{2}(\bar\chi^c\chi + \text{h.c.}), 
\end{equation}
where $\Delta m \ll m_\chi$. We predict that the Dirac fermion $\chi$ acquires a Majorana mass term, splitting into two separate mass eigenstates:
\begin{equation}
    \mathcal{L}_{\text{kinetic}}=\bar\chi_1(\mathrm i\slashed\partial-m_1)\chi_1+\bar\chi_2(\mathrm i\slashed\partial-m_2)\chi_2.
\end{equation}
In this way, the matrix is diagonalized and two Majorana fermions $\chi_1$ and $\chi_2$ are of masses
\begin{equation}
    m_1 = m_\chi - \frac{\Delta m}{2}, \qquad m_2 = m_\chi + \frac{\Delta m}{2},
\end{equation}
respectively. We regard $\chi_1$ as the stable dark matter candidate and $\chi_2$ as the heavier excited state.

\subsection{Symmetry and self-interaction}

Suppression of elastic self-scattering ($\chi_1 \chi_1 \rightarrow \chi_1 \chi_1$) at low velocities is a very powerful phenomenological necessity when addressing the tension between satellite survival and dwarf galaxy cores~\cite{Vogelsberger:2012ku}. If elastic and inelastic scatterings are equivalent, $v^{-4}$ cross-section in satellites would be expected to increase, which violates the Draco constraint~\cite{ODonnell:2025pkw}.

To block elastic scattering, we impose a dark $\mathbb{Z}_2$ parity. The fields then undergo the following transformations:
\begin{align}
    \chi_1 &\rightarrow +\chi_1 \quad (\text{even}), \\
    \chi_2 &\rightarrow -\chi_2 \quad (\text{odd}), \\
    \phi   &\rightarrow -\phi   \quad (\text{odd}).
\end{align}
Under this antisymmetric symmetry, a general renormalizable Yukawa-type interaction between the scalar and fermions is strictly off-diagonal:
\begin{equation}
    \mathcal{L}_{\text{int}} = -g_S \phi \left( \overline{\chi}_1 \chi_2 + \overline{\chi}_2 \chi_1 \right).
    \label{eq:scalar_lagrangian}
\end{equation}
Note that while the $\mathbb{Z}_2$ symmetry forbids the tree-level elastic vertex $\bar{\chi}_1 \chi_1 \phi$, residual elastic scattering is induced at one loop via box diagrams involving two scalar exchanges. However, these contributions are negligible, as demonstrated in Appendix~\ref{app:scattering}, and therefore do not violate stringent self-interaction bounds from satellite galaxies~\cite{ODonnell:2025pkw}.

This Lagrangian guarantees that the vertices $\chi_1\chi_1\phi$ and $\chi_2\chi_2\phi$ vanish identically. For $\chi_1\chi_2\rightarrow \chi_1\chi_2$, our derivation in Sec.~\ref{sec:cosmo} shows that the abundance of the excited state $\chi_2$ is effectively small, making tree-level elastic scattering negligible. We will see that tree-level $\chi_1 \chi_1 \rightarrow \chi_2 \chi_2$ scattering (or $\chi_1 \chi_2 \rightarrow \chi_2 \chi_1$) provides the desired phenomenology, while other processes are highly suppressed, if not forbidden.
While we treat the strict off-diagonal nature of the scalar interaction as an effective low-energy feature, its UV-complete origin requires careful construction. If the 20 MeV scalar mediator $\phi$ were simply the CP-even radial mode of a dark Higgs field that spontaneously breaks a dark $U(1)_D$ gauge symmetry, its couplings to the mass eigenstates would be strictly diagonal ($\phi \overline{\chi}_1 \chi_1$), as demonstrated in~\cite{Garcia-Cely:2013nin}. A light radial mode would therefore induce unsuppressed tree-level elastic scattering, violating the Draco constraints. Furthermore, the coupling of such a radial mode is fixed by $g_S \propto \Delta m / m_\phi$, which for our benchmark parameters would require a wildly non-perturbative quartic coupling $\lambda_\phi \sim 10^{10}$. To rigorously forbid these diagonal terms and decouple the interaction strength from the mass splitting, we separate the symmetry-breaking mechanism from the light SIDM mediator. Following the framework outlined in~\cite{Berlin:2025fwx}, we postulate a dark sector containing a heavy dark Higgs $\Phi_H$ and a separate light real scalar singlet $\phi$. The $U(1)_D$ gauge symmetry is broken at a high scale by $\langle \Phi_H \rangle$, generating the small Majorana mass splitting $\Delta m \approx 100$ eV. The dangerous diagonal elastic scattering mediated by the radial mode of $\Phi_H$ is heavily suppressed by its large mass ($m_{\Phi_H} \gg m_\chi$). The light mediator $\phi$, which does not acquire a vacuum expectation value, is instead governed by a dark charge-conjugation symmetry $\mathcal{C}_\chi$ under which the dark matter gauge eigenstates transform as $\chi \leftrightarrow \chi^c$. The mass eigenstates therefore transform with opposite $\mathcal{C}_\chi$ parities: $\chi_1 \to -\chi_1$ (odd) and $\chi_2 \to +\chi_2$ (even). By assigning $\phi$ to be explicitly odd under this symmetry ($\phi \to -\phi$), the diagonal interactions $\phi \overline{\chi}_1 \chi_1$ and $\phi \overline{\chi}_2 \chi_2$ are strictly forbidden. The parity-even Lagrangian then uniquely permits the interaction $\mathcal{L}_{\text{int}} = -g_S \phi (\overline{\chi}_1 \chi_2 + \text{h.c.})$. Because $\phi$ is independent of the symmetry-breaking VEV, its coupling $g_S$ is a free parameter, smoothly accommodating the strongly coupled resonant regime required for SIDM. This symmetry structure protects the kinematic-switch mechanism from dangerous diagonal corrections at tree level.
%While we treat the $\mathbb{Z}_2$ symmetry as an effective feature at low energies, it actually arises naturally from a UV-complete gauge structure. For instance, if the dark sector is charged under a $U(1)_D$ gauge symmetry, spontaneously broken by a dark Higgs field $S$ with charge $q_S=2$, the condensation of the Higgs ($\langle S \rangle \neq 0$) breaks $U(1)_D \rightarrow \mathbb{Z}_2$. This residual discrete symmetry guarantees the stability of the dark matter ground state and enforces the off-diagonal nature of the interactions, protecting the ``kinematic switch'' mechanism from dangerous diagonal corrections. 

\begin{figure}[htbp]
    \centering
    %plots should be of same height to calibrate the scale, 11->22 plot is 1:1 so we use them as standard
    \includegraphics[height=0.43\linewidth]{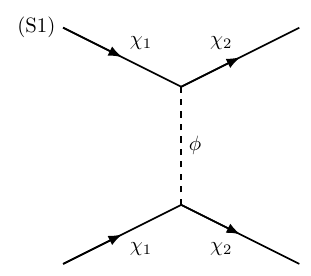}
    \hfill
    \includegraphics[height=0.43\linewidth]{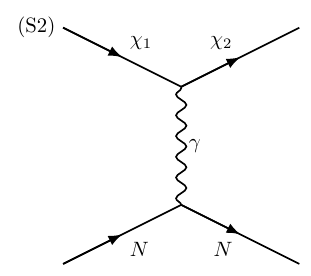}

    \includegraphics[height=0.43\linewidth]{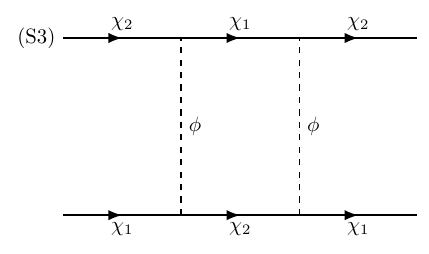}
    
    \includegraphics[height=0.43\linewidth]{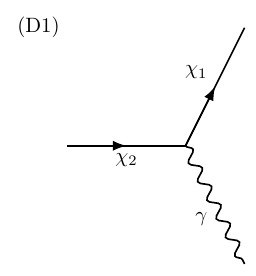}
    \hspace{.3cm}
    \includegraphics[height=0.43\linewidth]{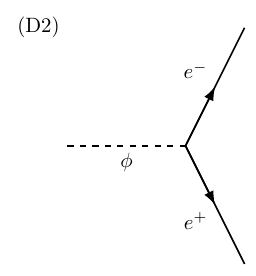}
    
    \caption{Feynman diagrams presenting some of the scatterings and decay processes to investigate: $\chi_1\chi_1\rightarrow\chi_2\chi_2$, the contributor to the inelastic scattering phenomenology (S1); $\chi_1N\rightarrow\chi_2N$, potential nuclear recoil in direct search experiments (S2); the possible elastic scattering below threshold (S3); $\chi_2\rightarrow\chi_1+\gamma$, the decay from excited state DM to stable state (D1); and $\phi\rightarrow e^++e^-$, decay of mediators to positron and electron pairs (D2). }
    \label{fig:Feynman_diagrams}
\end{figure}

\subsection{SM portal and decay of scalar mediator}

In the early universe, the dark sector was in thermal equilibrium with the plasma formed by SM particles. We presume the scalar $\phi$ is leptophilic. It couples to SM leptons ($l=e, \mu, \tau$) but not quarks or the Higgs boson. The effective Lagrangian for the portal is:
\begin{equation}
    \mathcal{L}_{portal} = \sum_{l} g_{l} \phi \overline{l} l,
    \label{eq:portal}
\end{equation}
where $g_{l}$ is the coupling constant. 

We treat the dark $\mathbb{Z}_2$ parity  as an approximate symmetry of the dark sector. While the dark sector fields transform non-trivially under this parity ($\chi_2 \to -\chi_2, \phi \to -\phi$), the Standard Model leptons are $\mathbb{Z}_2$-even.The portal interaction in Eq.~(\ref{eq:portal}) constitutes a source of explicit $\mathbb{Z}_2$ symmetry breaking. However, since the portal coupling is extremely weak ($g_e \sim 10^{-6}$) to satisfy beam-dump and stellar constraints, this breaking is parametrically small. We quantify the induced diagonal couplings and their impact on the stability of the kinematic switch in the following section.

The scalar mediator $\phi$ decays preferentially into electron-positron pairs ($e^+ e^-$)~\cite{Knapen:2017xzo}. When the mass of scalar mediator ranges from $10$ to $100$ MeV, the decay width is:
\begin{equation}
    \Gamma(\phi \rightarrow e^+ e^-) = \frac{g_e^2}{8\pi} m_\phi \left(1 - \frac{4 m_e^2}{m_\phi^2} \right)^{3/2}.
    \label{eq:decay_width_phi_pm}
\end{equation}
This leptophilic composition allows the mediator to rapidly decay prior to BBN by $\phi \rightarrow e^+e^-$ and falling within cosmological limits whilst simultaneously suppressing coupling to nucleons to avoid the limits introduced by direct search experiments~\cite{LZ:2022lsv, XENON:2025vwd, PandaX:2024qfu}, it gave $m_\chi$ from $10^{-2}$ to $10^2$ GeV. 

\subsection{The electron-specific portal and UV realization}
Throughout this work, we treat the electron portal coupling constant as a scan parameter in the range
\begin{equation}
g_e \in [10^{-7},\,10^{-5}],
\label{eq:ge_range}
\end{equation}
chosen to ensure mediator decay well before BBN. Also, it enables a conservative discussion of stellar/supernova constraints without committing to a single finely tuned value. Unless stated otherwise, numerical examples use reference value $g_e=10^{-6}$. In contrast, the dipole operator introduced in Sec.~\ref{sec:dipole} is controlled by an independent effective scale $\Lambda_{\rm eff}$ generated at loop level in a separate heavy charged sector and is not fixed by $g_e$.

Considering contributions to $(g-2)_\mu$ or $(g-2)_\tau$ factors, if the scalar mediator couples to the particles, the $(g-2)$ value will significantly deviate from SM prediction. Therefore, we make the assumption that $\phi$ only couples to electrons. We then construct a flavor-specific UV completion. 

Rather than a type-IV two-Higgs Doublet Model (2HDM)~\cite{Branco:2011iw}, we foresee the heavy vector-like leptons $E$ should couple to both $\phi$ and $e$ to make the coupling flavor specific. Our effective Lagrangian, below the mass scale $M_E$, is:
\begin{equation}
    \mathcal{L}_{\text{portal}} = g_e \phi \bar{e} e, \quad \text{with}\quad g_e \sim \lambda_\phi \frac{v_\text{EW}}{M_E} \sin\theta_{e}.
\end{equation}
We will discuss in Sec.~\ref{sec:dipole} that the transition magnetic dipole operator $\bar{\chi}_1 \sigma^{\mu\nu} \chi_2 F_{\mu\nu}/\Lambda$ is required to have a very high $\Lambda$ ($\sim 10^4$ GeV) in order to facilitate excited state decay. We note that generating this operator via loops involving the same weak portal coupling $g_e \approx 10^{-6}$ would imply a effective $\Lambda_\text{eff} \sim (16\pi^2 M)/(e g_S g_e) \gg 10^4$ GeV, which is phenomenologically insufficient. 
We therefore postulate a split-sector UV completion in which the portal coupling $g_e$ and the dipole operator are
generated by different heavy states.

We parametrize the dipole operator as an EFT generated at one loop by electrically charged
heavy states $\Psi$ of mass
\begin{equation}
M_\Psi \;=\; 10^{4}\ \mathrm{GeV}\,.
\label{eq:MPsi_def}
\end{equation}
At energies $\mu \ll M_\Psi$, integrating out $\Psi$ yields the transition dipole operator with a loop-suppressed
coefficient,
\begin{equation}
\mathcal L_{\rm dipole} \supset \frac{1}{\Lambda_{\rm eff}}\,
\bar\chi_1\sigma^{\mu\nu}\chi_2\,F_{\mu\nu} + \mathrm{h.c.},
\frac{1}{\Lambda_{\rm eff}}
\simeq
\frac{e\,\lambda_\Psi}{16\pi^{2}\,M_\Psi}\times \mathcal O(1)\,,
\label{eq:Lambdaeff_match}
\end{equation}
where $\lambda_\Psi$ denotes a representative product of dark-sector couplings in the UV completion and
$\mathcal O(1)$ encodes loop-function dependence on mediator mass ratios and chiral structure.
For $\lambda_\Psi=\mathcal O(1)$ and $M_\Psi=10^{4}\,\mathrm{GeV}$, this gives parametrically
$\Lambda_{\rm eff}\sim \mathrm{few}\times 10^{6}\,\mathrm{GeV}$.
Throughout the phenomenological analysis, we treat $\Lambda_{\rm eff}$ as the effective suppression scale
controlling both $\chi_2\to\chi_1\gamma$ and dipole-mediated scattering rates.
To generate the effective dipole scale $\Lambda_{eff} \sim 10^7$ GeV, we introduce a heavy sector of vector-like leptons $\Psi^{\pm}$ with mass $M_{\Psi} \sim 10-50$ TeV. These states carry dark charges and couple via $\mathcal{L}_{UV} \supset y \phi \bar{\Psi} \Psi + h.c.$ Integrating out $\Psi$ at one-loop generates the dipole operator with coefficient $\Lambda_{eff}^{-1} \sim e y^2 / (16\pi^2 M_{\Psi})$. This naturally accommodates the PeV-scale suppression.

\subsection{Dark dipole and excited state decay}
\label{sec:dipole}

For our benchmark parameters ($\Delta m \sim 100~\mathrm{eV}$, $m_\phi \sim 20~\mathrm{MeV}$), the decay
$\chi_2 \to \chi_1 \phi$ is kinematically forbidden $\Delta m \ll m_\phi$, so the excited state cannot
de-excite by emitting an on-shell mediator. We therefore include a dimension-5 transition electromagnetic
operator that permits radiative de-excitation,
$\chi_2 \to \chi_1 \gamma$. Importantly, the mediator lifetime relevant for BBN is controlled by the
leptophilic portal through $\phi\to e^+e^-$, which is prompt for our benchmark $g_e$.
The dipole operator instead controls the late-time fate of the excited state and the inelastic
photon-mediated scattering phenomenology discussed in Sec\ref{sec:scattering}. To stop the excited state $\chi_2$ from contributing too much to the relic and to allow stable decay to the SM states we introduce a dimension-5 transition magnetic dipole operator~\cite{Sigurdson:2004zp}, which links the dark sector to the photon field strength tensor $F_{\mu\nu}$:
\begin{equation}
    \mathcal L_{\rm dipole}=
\frac{1}{\Lambda_{\rm eff}}\,
\bar\chi_1\sigma^{\mu\nu}\chi_2F_{\mu\nu} + \mathrm{h.c.}
    \label{eq:dipole_lagrangian}
\end{equation}
Introducing this operator allows the decay $\chi_2 \rightarrow \chi_1 + \gamma$ to occur rapidly, giving a decay rate:
\begin{equation}
    \Gamma(\chi_2 \to \chi_1\gamma) \simeq \frac{1}{\pi}\,\frac{\Delta m^{3}}{\Lambda_{\rm eff}^{2}}\,
\end{equation}
For our benchmark values $\Lambda_{\rm eff}$ and $\Delta m \approx 100~\mathrm{eV}$, the excited-state lifetime
is $\tau_{\chi_2} = 1/\Gamma(\chi_2\to\chi_1\gamma)$.
For the direct-detection-safe parameter space of interest we typically have $\Lambda_{\rm eff}\gtrsim 10^{7}\,\mathrm{GeV}$,
so $\chi_2$ can be long-lived. This does not endanger BBN\cite{Depta:2020wmr} because the radiated energy per decay is only $\Delta m \ll \mathrm{MeV}$,
far below nuclear photodissociation thresholds; the relevant late-time limits are instead from electromagnetic energy injection such as spectral distortions\cite{Sigurdson:2004zp} and ionization history, which are strongly suppressed by the small fractional energy release
$\Delta m/m_\chi \ll 1$.

\section{Scattering Phenomenology}
\label{sec:scattering}

To solve the small-scale structure problems ($m_\chi \sim 10\text{--}100$ GeV, $m_\phi \sim 10$ MeV), the coupling strength $\alpha \equiv g_S^2/4\pi$ is typically large enough that the perturbative Born approximation fails. This failure occurs when the potential is strong enough to support bound states or resonances, formally when $\alpha m_\chi / m_\phi \gtrsim 1$. Our theory is in the non-perturbative ``resonant'' regime, where the cross-section exhibits a rich velocity dependence  defined as Sommerfeld enhancement~\cite{ArkaniHamed2009}. 

\subsection{Non-Perturbative Dynamics}

\begin{figure}[htbp]
    \centering
    \includegraphics[width=0.95\linewidth]{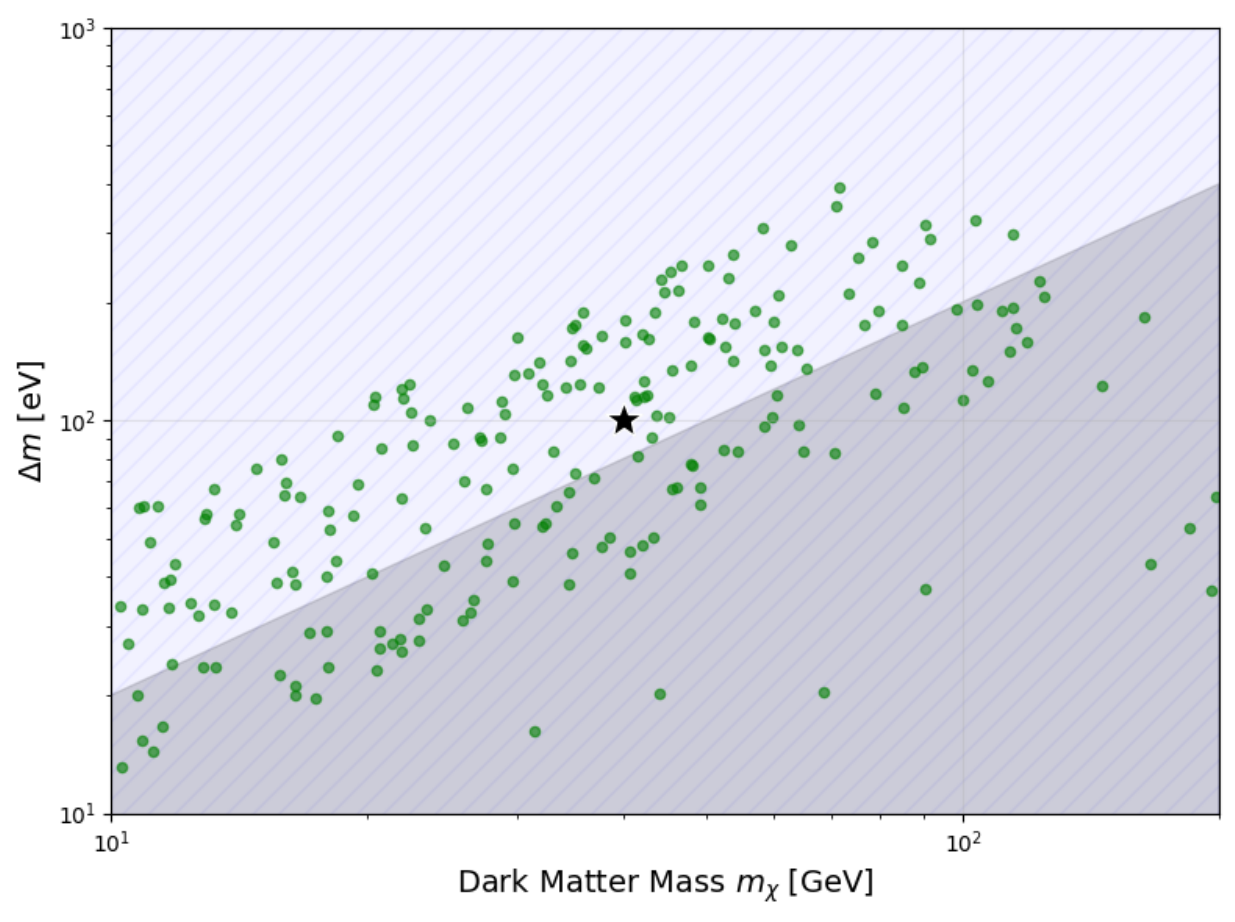}
    \caption{The discovery window. Green points represent the viable SIDM parameter space. The gray region is excluded by satellite stability. The benchmark point (star) sits within the discovery window for future low-threshold experiments via the dipole portal. }
    \label{fig:discovery}
\end{figure}

To compute the cross-section $\frac{\sigma_{T1\rightarrow2}}{m}$of inelastic scattering $\chi_1\chi_1\rightarrow\chi_2\chi_2$, we solve the non-relativistic coupled-channel Schr\"{o}dinger equation (see Appendix~\ref{app:scattering}). The system involves two states: the incoming $\chi_1 \chi_1$ pair and the excited $\chi_2 \chi_2$ pair. The off-diagonal interaction in Eq.~\eqref{eq:scalar_lagrangian} induces transitions between these states mediated by the Yukawa potential:
\begin{equation}
    V(r) = - \alpha \frac{\mathrm e^{-m_\phi r}}{r}.
\end{equation}
The inelastic channel $\chi_1 \chi_1 \rightarrow \chi_2 \chi_2$ only opens when the kinetic energy exceeds the mass splitting threshold, $E_\text{kin} > 2\Delta m$. Below this kinematic threshold, the dominant heat-transferring inelastic channel is closed. 
We note that the elastic scattering cross section for $\chi_1 \chi_1 \to \chi_1 \chi_1$ (S3 in Fig.~\ref{fig:discovery}) does not vanish exactly: although it is forbidden at tree level by the $\mathbb{Z}_2$ symmetry, it is induced by virtual transitions through the closed $\chi_2\chi_2$ channel (equivalently, by box-type contributions at $\mathcal{O}(\alpha^4)$ in the perturbative regime). In the resonant regime, however, the Born expansion can fail and the below-threshold elastic amplitude can be non-perturbatively enhanced by virtual closed-channel effects \cite{Schutz2015}. We therefore evaluate $\sigma(\chi_1\chi_1\to\chi_1\chi_1)$ non-perturbatively by solving the coupled-channel Schr\"odinger equation with an exponentially decaying boundary condition in the closed channel (Appendix~A.5), and select benchmarks that lie in an anti-resonance valley. For our benchmark, we obtain $\sigma/m_\chi \lesssim 0.01~\mathrm{cm}^2/\mathrm{g}$ at satellite velocities, safely below the Draco bound $\sigma/m_\chi \lesssim 0.1~\mathrm{cm}^2/\mathrm{g}$, thereby realizing the kinematic switch.
%We note that the elastic scattering cross-section $\chi_1 \chi_1 \rightarrow \chi_1 \chi_1$ (S3 in Fig.~\ref{fig:discovery}) does not vanish exactly: it persists via virtual transitions through the closed channel (equivalent to box diagrams at order $\alpha^4$). However, this residual elastic scattering is suppressed by factors of $\mathcal{O}(\alpha^2)$ relative to the resonant inelastic cross-section. In our benchmark regime, this residual component yields $\sigma < 0.01$ cm$^2$/g, which is well below the Draco constraint ($\sigma < 0.1$ cm$^2$/g), effectively realizing the kinematic switch.

\subsection{Resonant enhancement}

Above the kinematic threshold, the cross-section does not simply rise monotonically. It is heavily influenced by the presence of ``darkonium'' bound states in the spectrum~\cite{An:2016gad}. When the set of parameters $\{m_\chi, m_\phi, \alpha\}$ are tuned such that a bound state exists near zero energy, the scattering cross-section undergoes a resonant enhancement (the Sommerfeld effect). This resonance manifest as a sharp peak in the cross-section just above the threshold velocity\cite{Hisano:2003ec}.

For our benchmark parameter ($m_\chi \approx 40$ GeV), we reside near a resonance. This leads to the following behavior:

\begin{itemize}
    \item[(a)] Sub-threshold regime ($v < v_\text{th}$): In ultra-faint satellites ($v \sim 20$ km/s), the kinetic energy is insufficient to create the heavier $\chi_2$ state~\cite{Ando:2025qtz}. Inelastic scattering is forbidden, and residual elastic scattering is negligible, preventing core collapse. 
    \item[(b)] Resonant regime ($v > v_\text{th}$): The channel opens in dwarf galaxies ($v \sim 60$ km/s). The proximity to the resonance pole amplifies the cross-section by orders of magnitude compared to the geometric size of the target, reaching $\sigma/m \sim 10$ cm$^2$/g. This is the mechanism that generates the large cores observed in LSB galaxies~\cite{Correa:2020qam}. 
    \item[(c)] Classical regime ($v \gg v_\text{th}$): At cluster velocities ($v \sim 10^3$ km/s), the kinetic energy is much larger than the binding energy of the resonance~\cite{Tulin2013}. The Sommerfeld enhancement is negligible~\cite{ArkaniHamed2009}, and the cross-section falls off according to the classical Rutherford scaling $\sigma \propto v^{-4}$.
\end{itemize}

\section{Cosmological Constraints}
\label{sec:cosmo}

\subsection{Relic density and thermal freeze-out}
\label{sec:relic_main}

Since the mass splitting is tiny relative to DM mass states, $\Delta m \ll T_f \simeq m_\chi/20$, the excited state remains in chemical equilibrium with the ground state during freeze-out, which gives $n_1^{\rm eq}\simeq n_2^{\rm eq}\simeq \tfrac12 n_{\rm tot}^{\rm eq}$. The off-diagonal $\mathbb Z_2$-protected interaction in Eq.~\eqref{eq:scalar_lagrangian} forbids the mixed annihilation channel $\chi_1\chi_2\rightarrow\phi\phi$ at tree level, while $\chi_1\chi_1\rightarrow\phi\phi$ and $\chi_2\chi_2\rightarrow\phi\phi$ are allowed and equal up to $\mathcal{O}(\Delta m/m_\chi)$ corrections. The effective annihilation rate entering the single Boltzmann equation for the total density is
\begin{equation}
    \label{eq:sigeff_main}
    \langle\sigma v\rangle_{\rm eff}(T)
    =\sum_{i,j=1,2}\frac{n_i^{\rm eq}n_j^{\rm eq}}{(n_{\rm tot}^{\rm eq})^2}\langle\sigma v\rangle_{ij}(T)
    \simeq \frac12\,\langle\sigma v\rangle_{11}(T).
\end{equation}

The short-distance annihilation process $\chi_1\chi_1\rightarrow\phi\phi$ is $p$-wave suppressed in the non-relativistic limit~\cite{Tulin2013}, $(\sigma v)_{\rm Born}(v)=\sigma_0 v^2$ with $\sigma_0\equiv 3\pi\alpha^2/(2m_\chi^2)$. Long-range effects from the Yukawa force must be treated non-perturbatively in our resonant parameter region $\alpha m_\chi/m_\phi \gtrsim 1$. We therefore include a coupled-channel Sommerfeld enhancement factor $S_p(v)$ computed from the same potential parameters $(\alpha,m_\phi,\Delta m)$ as used in the scattering calculation (Appendix~\ref{app:scattering}). The Sommerfeld-enhanced and thermally averaged rate is
\begin{equation}
    \label{eq:thermal_main}
    \langle\sigma v\rangle_{11}(x)=\frac{x^{3/2}}{2\sqrt{\pi}}\int_0^\infty \d v\,v^2\,\mathrm e^{-x v^2/4}\,(\sigma v)_{\rm Born}(v)\,S_p(v),
\end{equation}
where $x\equiv {m_\chi/T}$. We solve the standard freeze-out Boltzmann equation using $\langle\sigma v\rangle_{\rm eff}(x)=\tfrac12\langle\sigma v\rangle_{11}(x)$.
The full thermal-averaging procedure and the consistent computation of $S_p(v)$ from the coupled-channel Schr\"odinger
equation are given in Appendix~\ref{app:relic}.
To ensure physical consistency in the resonant regime, we must account for the unitarity limit of the p-wave annihilation cross-section. The maximum possible inelastic cross-section for the $l=1$ partial wave is strictly bounded by the unitarity of the S-matrix, scaling as $\sigma_{uni}^{(l=1)} \approx 3\pi/k_{rel}^2$. Standard factorization, $\sigma \propto (\sigma v)_{Born} S_p(v)$, can violate this bound near narrow resonances where the enhancement is large. A fully rigorous treatment involves determining the annihilation rate from the imaginary part of the optical potential in the S-matrix formalism\cite{Parikh:2024mwa} , such a computation is computationally prohibitive for a global parameter scan.

Instead, we adopt a unitarized prescription~\cite{Blum:2016nrz} that enforces saturation smoothly:
\begin{equation}
    \sigma_{11}^{unit}(v) = \frac{(\sigma v)_{Born} S_p(v)/v}{1 + \frac{(\sigma v)_{Born} S_p(v)/v}{\sigma_{uni}^{(l=1)}}} \simeq \frac{(\sigma v)_{Born} S_p(v)}{v \left(1 + \frac{(\sigma v)_{Born} S_p(v) k_{rel}^2}{3\pi v} \right)}.
    \label{eq:unitarity}
\end{equation}
This expression reduces to the standard Sommerfeld-enhanced rate in the perturbative regime but saturates at the unitarity bound exactly on resonance. We implement this unitarized rate in the thermal average integral  for the relic density calculation. We explicitly verified for our benchmark parameters ($\alpha \approx 10^{-2}$), the factorized cross-section approaches the unitarity limit only in a extremely narrow velocity window near the resonance peak. Since the relic density is determined by a broad thermal average ($x_f \sim 20$), the precise shape of the cutoff at the pole has a negligible impact ($< 1\%$) on the final abundance compared to the Parikh-Sato-Slatyer limit.

Finally, Bound-State Formation (BSF) is negligible for our benchmark because radiative capture via on-shell mediator emission, $\chi\chi\rightarrow B+\phi$, is kinematically forbidden when $m_\phi>E_B$. Also, the leading off-shell process $\chi\chi\rightarrow B+\phi^\ast\rightarrow B+e^+e^-$ is suppressed by both the heavy propagator and the small leptophilic portal coupling. The quantitative estimate is given in Appendix~\ref{app:relic} as well.

\subsection{CMB anisotropies and p-wave suppression}
Dark matter annihilation during the recombination epoch ($z \sim 1100$) injects energy into the primordial plasma, dampening the temperature anisotropies of the CMB~\cite{Slatyer:2015jla}. Planck～\cite{Planck2018} constrains late-time electromagnetic energy injection through the parameter
\begin{equation}
    \label{eq:pann_def}
    p_{\rm ann}\;\equiv\; f_{\rm eff}\,\frac{\langle\sigma v\rangle}{m_\chi},
\end{equation}
where $f_{\rm eff}$ is the energy-deposition efficiency (channel- and mass-dependent) evaluated at $z\simeq 600$～\cite{Slatyer:2015jla}. For $s$-wave annihilation Planck~\cite{Planck2018} gives the approximate bound
\begin{equation}
    \label{eq:planck_bound}
    p_{\rm ann} \;<\; 3.2\times 10^{-28}\;{\rm cm^3\,s^{-1}\,GeV^{-1}}
    \qquad (95\%~{\rm C.L.})\;,
\end{equation}
which we use as a conservative reference constraint. For velocity-suppressed annihilation the true constraint is weaker than the $s$-wave limits; using Eq.~\eqref{eq:planck_bound} is therefore conservative for our $p$-wave model. 

In our model the primary annihilation products are $\phi\phi$, and each $\phi$ promptly decays to an $e^+e^-$ pair. Thus the deposited power is that of an electromagnetic cascade initiated by four relativistic leptons. We incorporate this using an effective efficiency $f_{\rm eff}^{\phi\phi\rightarrow 4e}(m_\chi,m_\phi)$. For $m_\phi\ll m_\chi$, each lepton carries energy $\sim m_\chi/2$ and we adopt
\begin{equation}
    \label{eq:feff_choice}
    f_{\rm eff}\;\equiv\; f_{\rm eff}^{e^\pm}(E\simeq m_\chi/2)\,,
\end{equation}
using the tabulated energy-dependent efficiencies from Ref.~\cite{Slatyer:2015jla} (evaluated at $z\simeq 600$). Numerically, for $m_\chi\simeq 40~{\rm GeV}$ this corresponds to $f_{\rm eff}\sim \mathcal{O}(0.3)$ for $e^+e^-$-initiated cascades.

The relevant annihilation parameter is therefore
\begin{equation}
    \label{eq:pann_model}
    p_{\rm ann}^{\rm model}
    \;=\;
    f_{\rm eff}\,\frac{\langle\sigma v\rangle_{\rm eff}(z\simeq 600)}{m_\chi}
    \;\simeq\;
    f_{\rm eff}\,\frac{1}{2}\,\frac{\sigma_0\,v_{\rm rec}^2\,S_{p,{\rm max}}}{m_\chi}\,,
\end{equation}
with $v_{600}\sim 10^{-7}$ the typical DM velocity around $z\simeq 600$ and $S_{p,{\rm max}}$ the saturated Sommerfeld factor. For our benchmark the resulting $p_{\rm ann}^{\rm model}$ is many orders of magnitude below the Planck limit in Eq.~\eqref{eq:planck_bound}, even taking $f_{\rm eff}=\mathcal{O}(1)$. 

\subsection{CMB safety and kinematic blocking of bound states}
Dark matter annihilation during the recombination epoch ($z \sim 1100$) is strictly constrained by Planck~\cite{Planck2018}. Our model evades these constraints via two mechanisms: $p$-wave suppression of direct annihilation and the kinematic blocking of bound-state formation (BSF).

\paragraph{Direct annihilation:}
The process $\chi_1 \chi_1 \rightarrow \phi \phi$ is $p$-wave suppressed for Majorana fermions annihilating into scalars. At recombination velocities ($v \sim 10^{-7}$), the rate scales as $\langle \sigma v \rangle \propto v^2$, rendering energy injection negligible.

\paragraph{BSF:}
For our benchmark the mediator is heavier than the would-be binding energy, so radiative BSF via on-shell scalar emission, $\chi\chi\rightarrow \mathcal{B}+\phi$, is kinematically forbidden when $m_\phi$ is larger than the bound state energy $E_B$. A representative Coulombic estimate~\cite{Petraki:2016cnz} for the ground-state binding energy is
\begin{equation}
    E_B \sim \frac{\alpha^2 m_\chi}{4},
    \label{eq:EB_est}
\end{equation}
which gives $E_B\sim 1~{\rm MeV}$ for $\alpha=10^{-2}$ and $m_\chi=40~{\rm GeV}$, well below the limit $m_\phi=20~{\rm MeV}$.

The leading allowed capture channel proceeds through an off-shell scalar that subsequently decays, $\chi\chi\rightarrow \mathcal{B}+\phi^\ast\rightarrow \mathcal{B}+e^+e^-$. For momentum transfer $q^2\sim E_B^2\ll m_\phi^2$, the virtual propagator yields a strong suppression, $|(q^2-m_\phi^2)^{-1}|^2\simeq m_\phi^{-4}$, and the rate inherits an additional small-coupling penalty from $\phi^\ast\rightarrow e^+e^-$. A conservative estimate of the off-shell-to-on-shell capture-rate ratio is therefore
\begin{equation}
    \frac{\Gamma_{\rm off}}{\Gamma_{\rm on}}
    \sim \frac{\alpha_{\rm EM}\,g_e^2}{4\pi}\left(\frac{E_B}{m_\phi}\right)^4
    \times \mathcal{O}(1),
    \label{eq:BSF_offshell_scaling}
\end{equation}
where the factor $(E_B/m_\phi)^4$ encodes the squared propagator and the soft-emission kinematics. Numerically, with $E_B\simeq 1~{\rm MeV}$, $m_\phi=20~{\rm MeV}$, and $g_e=10^{-6}$, one has
\begin{equation}
    \frac{\Gamma_{\rm off}}{\Gamma_{\rm on}}\lesssim 10^{-21},
\end{equation}
Therefore, off-shell BSF is utterly negligible both during freeze-out and at recombination, and does not compete with the Sommerfeld-enhanced annihilation channels considered in the Boltzmann equation.

\subsection{Big bang nucleosynthesis}
The light scalar mediator $\phi$ must decay before the onset of BBN ($t \sim 1$ s), otherwise the element abundance in today's universe will not match the observations. Our leptophilic portal ensures this via the decay $\phi \rightarrow e^+ e^-$. For a coupling $g_e \sim 10^{-6}$ and $m_\phi = 20$ MeV:
\begin{equation}
    \tau_\phi \approx 0.8 \times 10^{-9} \, \text{s} \ll 1 \, \text{s}.
    \label{eq:phi_life_time}
\end{equation}
This rapid decay ensures that the dark sector entropy is transferred to the SM plasma well before neutrino decoupling. Unlike the Higgs portal~\cite{Knapen:2017xzo}, the leptophilic coupling does not require mixing with quarks, decoupling the BBN lifetime constraint from nuclear recoil constraints.

While on-shell mediator emission is forbidden, bound state formation can proceed via off-shell scalars decaying to electron-positron pairs: $\chi \chi \rightarrow \mathcal{B} + \phi^* \rightarrow \mathcal{B} + e^+ e^-$. This process is kinematically allowed only if the binding energy exceeds the electron-positron mass threshold, $E_B > 2m_e \approx 1.02$ MeV. For our benchmark $\alpha \approx 10^{-2}$ and $m_\chi = 40$ GeV, the binding energy is $E_B \approx \frac{1}{4}\alpha^2 m_\chi \approx 1.0$ MeV. This lies just below the $2m_e$ threshold, rendering the off-shell process kinematically blocked or heavily phase-space suppressed. We impose the constraint $\alpha \lesssim 0.01$ to ensure $E_B < 2m_e$, thereby maintaining CMB robustness against all BSF channels.

\subsection{Robustness of CMB constraints via saturation}
The Sommerfeld enhancement $S_p \propto v^{-2}$ must physically saturate, otherwise we will have an unphysical infinity cross section when DM particles cool ($v\rightarrow 0$) as it scales with $1/v^2$. We quantify the saturation explicitly. Saturation occurs when the particle's de Broglie wavelength is cut off by the finite range of the mediator, roughly at velocities $v_\text{sat} \sim m_{\phi}/m_{\chi}$. 

For our benchmark ($m_{\phi}=20$ MeV, $m_{\chi}=40$ GeV), $v_\text{sat} \approx 5 \times 10^{-4}c$. During recombination ($z \sim 1100$), the dark matter velocity is $v_\text{rec} \sim 10^{-7}c$, placing the system deep in the saturation regime ,$v_\text{rec} \ll v_\text{sat}$.
so $S_p(v)$ has saturated to
\begin{equation}
    S_{p,\max}\equiv S_p(v\ll v_{\rm sat})\simeq \left(\frac{v_\star}{v_{\rm sat}}\right)^2.
\end{equation}
The CMB-relevant rate therefore scales as
\begin{equation}
    \langle\sigma v\rangle_{\rm eff}(v_{\rm rec})
    \simeq \frac{1}{2}\,\sigma_0\,v_{\rm rec}^2\,S_{p,\max},
    \qquad
    \sigma_0=\frac{3\pi\alpha^2}{2m_\chi^2}.
    \label{eq:sigmav_rec_pwave_sat}
\end{equation}
For benchmark values $(m_\chi,m_\phi,\alpha)=(40~{\rm GeV}, 20~{\rm MeV}, 10^{-2})$, one has $\sigma_0\simeq 2.9\times10^{-7}~{\rm GeV}^{-2}$, i.e.
\begin{equation}
    \sigma_0 \simeq 3.4\times10^{-24}\ {\rm cm}^3/{\rm s}.
\end{equation}
Hence we have
\begin{equation}
    \langle\sigma v\rangle_{\rm eff}(v_{\rm rec})
    \simeq 1.7\times10^{-38}\ {\rm cm}^3/{\rm s}\times S_{p,\max},
\end{equation}
which remains negligible compared to the Planck bound even for very large saturated enhancements. This explicitly shows that saturation does not remove the $v^2$ suppression at recombination; it only replaces the $v^{-2}$ growth of $S_p$ by a constant factor.

\subsection{Mediator lifetime and laboratory constraints}
To satisfy BBN requirements while evading beam-dump constraints, we adopt a benchmark coupling of $g_{e} = 10^{-6}$ and $m_{\phi} = 20$ MeV. The leptophilic scalar decays into electron-positron pairs with a decay width given by Eq.~\eqref{eq:decay_width_phi_pm}. For our benchmark, this yields a proper lifetime $\tau_{\phi} \approx 6 \times 10^{-10}$ s and a decay length $c\tau \approx 18$ cm. The lifetime is significantly shorter than the onset of BBN ($\tau_{\phi} \ll 1$ s). The dark sector entropy is transferred to the SM plasma well before neutrino decoupling ($T \sim 2$ MeV), ensuring $N_\text{eff}$ undisturbed.
With these results, the scalar is long-lived on the scale of collider vertex detectors but decays promptly relative to beam-dump baselines. This parameter space falls into a known gap in experimental coverage.For Beam Dumps experiments sucha s E137, they typically require long-lived particles to penetrate meters of shielding.The decay length of $c\tau \approx 18$ cm is thus the scalar decays primarily within the shield. The E137 constraints exclude $g_e \lesssim 10^{-7}$ for longer lifetimes\cite{Knapen:2017xzo,Beacham:2019nyx}. Searches for visible dark photons or scalars at BaBar ($e^+ e^- \to \gamma \phi$) and NA64 ($eN \to eN \phi$) currently constrain couplings $g_e \gtrsim 10^{-4}$ in this mass range~\cite{BaBar:2014zli, Banerjee:2019pds}. Our benchmark coupling of $g_e \approx 10^{-6}$ is two orders of magnitude below these limits, primarily due to the p-wave suppression of the scalar production cross-section relative to vector models.

However, this viable window is the primary target for next-generation missing-momentum experiments. The LDMX experiment, designed to search for light dark matter via bremsstrahlung production ($e Z \to e Z \phi$), projects sensitivity to scalar couplings down to $g_e \sim 10^{-6}$ in the $10-100$ MeV range~\cite{LDMX:2018cma}. A signal in LDMX, potentially distinguishable via the kinematics of the prompt decay, would serve as a complementary terrestrial probe to the direct detection.

\subsection{Kinetic Decoupling and Drag}
Scattering between dark matter and leptons ($\chi e \rightarrow \chi e$) maintains kinetic equilibrium in the early universe. For our benchmark parameters ($g_e \approx 10^{-6} ,m_\phi = 20$ MeV), the momentum transfer rate $\Gamma_\text{kd}$ drops below the Hubble rate $H$ before recombination. From standard WIMP decoupling, the decoupling temperature is roughly $T_\text{kd} \sim 10$ MeV~\cite{Binder:2016pnr}. Since $T_\text{kd} \gg T_\text{rec} \sim 0.3$ eV, the dark matter is kinetically decoupled during the formation of the CMB and Large Scale Structure. Therefore there is no significant DM-lepton drag to suppress the matter power spectrum or alter CMB acoustic peaks~\cite{Boehm:2001hm}. We verify the kinetic decoupling temperature $T_{kd}$ is determined by equating the momentum transfer rate $\gamma(T)$ to the Hubble rate $H(T)$. For $t$-channel scattering $\chi e \to \chi e$:
\begin{equation}
\gamma(T) \approx \frac{31\pi^3}{9} \frac{g_S^2 g_e^2 T^6}{m_{\chi} m_{\phi}^4}.
\end{equation}
Solving for our benchmark yields $T_{kd} \approx 15$ MeV. Since $T_{kd} \gg T_{BBN} \sim 1$ MeV, the dark sector is fully decoupled during nucleosynthesis.
Comparing this to $H \sim T^2/M_\text{Pl}$, and using $g_e = 10^{-6}$, $\alpha = 10^{-2}$, we find that kinetic decoupling occurs at $T_\text{kd} \sim 10-50$ MeV. Crucially, $T_\text{kd} \ll T_\text{freeze-out} (\approx 2 \text{ GeV})$, ensuring that the relic density calculation in the standard thermal bath remains valid. Furthermore, $T_\text{kd} \gg T_\text{BBN} (\sim 1 \text{ MeV})$ implies that the dark fluid is fully decoupled and cold during nucleosynthesis and structure formation.

\subsection{Supernova trapping}
Constraints from Supernova 1987A is significant.\cite{Chang:2018rso} It is driven by the Raffelt criterion\cite{Raffelt:1996wa} which forbids exotic energy loss rates exceeding $L_{\nu} \sim 3 \times 10^{52}$ erg/s. For dark sector it excludes a cooling wedge of couplings where particles are produced efficiently but escape the core. 
In the supernova core ($\rho \approx 3 \times 10^{14}$ g/cm$^3$, $T \approx 30$ MeV), the electron chemical potential $\mu_e \approx 300$ MeV significantly exceeds the scalar mass $m_\phi = 20$ MeV. The standard decay channel $\phi \to e^+ e^-$ is heavily Pauli-blocked, as the final-state electrons cannot access the occupied states below the Fermi surface ($E_e \sim m_\phi/2 \ll \mu_e$). Therefore, the trapping argument cannot rely on simple inverse decay\cite{Boehm:2003hm}.

Instead,the opacity is dominated by absorption via inverse bremsstrahlung on protons ($\phi e p \to e p$) and Compton-like scattering ($\phi e \to \gamma e$). Following the degenerate plasma formalism of Chang et al.~\cite{Chang:2018rso}, the absorptive width for a scalar in this regime is given by:
\begin{equation}
    \Gamma_{abs} \approx \sigma_{brem} n_p v_{rel} \approx \frac{4\pi \alpha_{EM} g_e^2 n_p}{3 m_e^2 E_\phi} \mathcal{F}_{deg},
\end{equation}
where $\mathcal{F}_{deg}$ represents the phase-space factors for the degenerate electrons. For our coupling $g_e = 10^{-6}$, the absorption mean free path is:
\begin{equation}
    \lambda_{abs} \equiv \Gamma_{abs}^{-1} \sim \mathcal{O}(10^2 \text{ cm}).
\end{equation}
this is negligible compared to the core radius ($R_{core} \sim 10$ km). The characteristic diffusion time for the scalar to escape is:
\begin{equation}
    \tau_{diff} \sim \frac{R_{core}^2}{c \lambda_{abs}} \sim 10^3 \text{ s}.
\end{equation}
Since $\tau_{diff} \gg \tau_{\nu} (\sim 10 \text{ s})$, the scalar fluid is fully thermalized and trapped within the neutrinosphere. It does not contribute to free-streaming cooling but participates in the thermal transport. Thus, the Pauli blocking of decays does not re-open the constraint window for $g_e \sim 10^{-6}$. 

\subsection{Excited state decay}
The electromagnetic energy injection from $\chi_2\rightarrow\chi_1\gamma$ depends on the excited-state fraction at the time of decay. After chemical decoupling, the comoving number density of $\chi_2$ simply redshifts and decays,
so the fraction evolves as
\begin{equation}
    \label{eq:chi2_fraction}
    f_{\chi_2}(t)\equiv \frac{n_{\chi_2}(t)}{n_{\rm DM}(t)} \simeq f_{\chi_2}(t_{\rm fo})\,\mathrm e^{-t/\tau_{\chi_2}}\,,
\end{equation}
where $f_{\chi_2}(t_{\rm fo})\simeq 1/2$ during freeze-out, and $\tau_{\chi_2}=1/\Gamma(\chi_2\rightarrow\chi_1\gamma)$ is given by Eq.~\eqref{eq:dipole_lagrangian}. 

The total electromagnetic energy released per decay is $\Delta m$. The fractional energy injected into the photon bath at time $t$ is therefore
\begin{equation}
    \label{eq:delta_rho_gamma}
    \frac{\Delta\rho_\gamma}{\rho_\gamma}\bigg|_{t}
    \simeq
    f_{\chi_2}(t)\,
    \frac{\rho_{\rm DM}(t)}{\rho_\gamma(t)}\,
    \frac{\Delta m}{m_\chi}\,,
\end{equation}
up to an $\mathcal{O}(1)$ deposition factor for sub-keV photons. Evaluating near the decay time $t\simeq \tau_{\chi_2}$ (where $f_{\chi_2}\simeq f_{\chi_2}(t_{\rm fo})/e$), and using $\Delta m=100~{\rm eV}$ and $m_\chi=40~{\rm GeV}$, we obtain
\begin{equation}
    \label{eq:delta_rho_num}
    \frac{\Delta\rho_\gamma}{\rho_\gamma}\bigg|_{t\simeq \tau_{\chi_2}}
    \ll 10^{-4}\,,
\end{equation}
well below the COBE/FIRAS bound on $\mu$- and $y$-type spectral distortions~\cite{Chluba:2011hw}.
Moreover, the emitted photons have energy $\Delta m\ll \mathcal O({\rm MeV})$ and therefore cannot photodissociate light nuclei, so there is no BBN photodissociation constraint. We emphasize that $p_{\rm ann}$ constraints apply to annihilation-like power injection near recombination\cite{Planck2018},
whereas $\chi_2$ decays are constrained by spectral distortions and other limits\cite{Poulin:2016anj} on late electromagnetic
energy injection, with the impact strongly suppressed by $\Delta m/m_\chi\ll 1$.

\subsection{Excited-state depletion and conversion freeze-out}
\label{sec:chi2_depletion}

A crucial assumption is that the excited-state fraction
$f_{\chi_2}\equiv n_{\chi_2}/(n_{\chi_1}+n_{\chi_2})$ becomes negligible at late
times. The ratio $n_{\chi_2}/n_{\chi_1}$ tracks its equilibrium value only
through number-changing conversion processes of interchanging $\chi_1$ and
$\chi_2$ that is efficient compared to Hubble expansion.

In the $Z_2$ , the dominant conversion reactions are
pair conversion and its inverse,
\begin{equation}
\chi_2\chi_2 \to \chi_1\chi_1,\qquad
\chi_1\chi_1 \to \chi_2\chi_2,
\end{equation}
Single-$\chi_2$ conversion processes such as $\chi_2\chi_1\to\chi_1\chi_1$
are forbidden by $Z_2$. In addition, the transition dipole operator
induces radiative de-excitation $\chi_2\to \chi_1\gamma$ with rate
$\Gamma_{\rm dec}\equiv \Gamma(\chi_2\to\chi_1\gamma)$.

After chemical freeze-out of the total comoving DM density
$n_{\rm DM}\equiv n_{\chi_1}+n_{\chi_2}$, the evolution of the excited fraction
$f\equiv f_{\chi_2}$ is governed by
\begin{equation}
\dot f
=
-2\,n_{\rm DM}\!\left[
\langle\sigma v\rangle_{22\to 11}\,f^2
-
\langle\sigma v\rangle_{11\to 22}\,(1-f)^2
\right]
-\Gamma_{\rm dec}\,f,
\label{eq:fdot_master}
\end{equation}
where dots denote derivatives with respect to cosmic time, and
$n_{\rm DM}\propto a^{-3}$. The up- and down-conversion rates are related by
detailed balance,
\begin{equation}
\langle\sigma v\rangle_{11\to 22}
=
\langle\sigma v\rangle_{22\to 11}\,
\left(\frac{n^{\rm eq}_{\chi_2}}{n^{\rm eq}_{\chi_1}}\right)^{\!2}
\simeq
\langle\sigma v\rangle_{22\to 11}\,e^{-2\Delta m/T_{\chi}},
\label{eq:detailed_balance}
\end{equation}
where $T_{\chi}\equiv T_{\rm dark}$ is the dark-matter kinetic temperature.

The conversion processes maintain $f\simeq f_{\rm eq}$ as long as the conversion
rate per excited particle exceeds Hubble. 
\begin{equation}
\Gamma_{\rm conv}(T)\;\equiv\; 2\,n_{\chi_2}\,\langle\sigma v\rangle_{22\to 11}
\;=\;2\,f\,n_{\rm DM}\,\langle\sigma v\rangle_{22\to 11}
\;\gtrsim\; H(T).
\label{eq:conv_condition}
\end{equation}
When $\Gamma_{\rm conv}\lesssim H$, the excited fraction freezes out at a
residual value and  redshifts.

For the cosmological inputs, we use
\begin{equation}
n_{\rm DM}(T)=\frac{\rho_{\rm DM}(T)}{m_\chi}
=\frac{\Omega_{\rm DM}\rho_c}{m_\chi}(1+z)^3,
H(T)=1.66\,\sqrt{g_*}\,\frac{T^2}{M_{\rm Pl}}.
\label{eq:nDM_and_H}
\end{equation}
After kinetic decoupling at temperature $T_{\rm kd}$, the DM temperature cools
as $T_{\chi}\propto a^{-2}$, implying (for $T<T_{\rm kd}$)
\begin{equation}
T_{\chi}(T)\simeq \frac{T^2}{T_{\rm kd}}.
\label{eq:Tchi_scaling}
\end{equation}
We determine $T_{\rm kd}$ same previously  and compute
$\langle\sigma v\rangle_{22\to 11}(T_{\chi})$ by thermally averaging the
velocity-dependent exothermic conversion cross section.\ref{app:scattering}

Solving Eq.~\eqref{eq:fdot_master} with
Eqs.~\eqref{eq:detailed_balance}--\eqref{eq:Tchi_scaling}, we find that
conversion remains efficient down to $T_{\chi}\lesssim \Delta m$.Then 
$f_{\chi_2}$ becomes exponentially small. For the benchmark
$\Delta m\simeq 100~\mathrm{eV}$ we obtain $f_{\chi_2}\ll 10^{-6}$ by
recombination, validating the approximation of a pure ground-state halo at late
times and rendering constraints from late $\chi_2$ decays negligible due to the
tiny excited fraction and the small fractional energy release $\Delta m/m_\chi$.
```latex
%========================
%  Section IV.C (NEW/UPDATED)
%========================
\subsection{Late-time EUV emission and molecular-cloud ionization constraints}
\label{sec:late-time-euv-mc}

In the Milky Way halo, the endothermic inelastic process
$\chi_1\chi_1\to \chi_2\chi_2$ is kinematically open at typical Galactic
velocities. Taking a one-dimensional velocity dispersion
$v_0\simeq 220~{\rm km/s}$, the characteristic relative speed is
$v_{\rm rel}\sim \sqrt{2}\,v_0\simeq 311~{\rm km/s}$, implying a typical
center-of-mass kinetic energy
\begin{equation}
E_{\rm kin} \;=\; \frac{\mu v_{\rm rel}^2}{2}
\;=\;\frac{m_\chi v_{\rm rel}^2}{4}
\;\sim\; {\cal O}(10~{\rm keV})
\;\gg\; 2\Delta m,
\end{equation}
so upscattering proceeds efficiently in the Galactic environment while
remaining closed in sufficiently cold systems such as dwarf galaxies.

Once produced, the excited state $\chi_2$ decays through the dipole operator,
$\chi_2\to\chi_1\gamma$, emitting a monoenergetic photon with
$E_\gamma\simeq \Delta m$. For the benchmark values used 
the decay rate is 
\begin{equation}
\Gamma_{\chi_2} \;\simeq\; \frac{1}{\pi}\,\frac{\Delta m^3}{\Lambda_{\rm eff}^2},
\qquad
\tau_{\chi_2}\equiv \Gamma_{\chi_2}^{-1}.
\label{eq:chi2-width}
\end{equation}
For $\Lambda_{\rm eff}=10^7~{\rm GeV}$ and $\Delta m=100~{\rm eV}$ this gives
$\tau_{\chi_2}\sim 2\times 10^{11}~{\rm s}$, corresponding to a decay length
$d\sim v_0\tau_{\chi_2}\sim {\cal O}({\rm pc})$, much smaller than Galactic-scale
density gradients. In this regime the excited state reaches local steady state.
 The local upscattering event rate is\cite{ONeil:2022szc}
\begin{equation}
R_{\uparrow}(\mathbf{x}) \;=\; \frac{1}{2}\,n_1^2(\mathbf{x})\,
\langle\sigma v\rangle_{11\to 22},
\end{equation}
each event produces two $\chi_2$ states and  each $\chi_2$ yields one photon upon decay. The local photon
emissivity is therefore
\begin{equation}
q_\gamma(\mathbf{x})
\;\simeq\; \Gamma_{\chi_2}\,n_2(\mathbf{x})
\;\simeq\; 2R_{\uparrow}(\mathbf{x})
\;=\; n_1^2(\mathbf{x})\,\langle\sigma v\rangle_{11\to 22}.
\label{eq:qgamma-local}
\end{equation}
We evaluate $\langle\sigma v\rangle_{11\to 22}$ by integrating our exact
non-perturbative inelastic cross section over the Milky Way velocity
distribution; for the benchmark parameters we find
$\langle\sigma v\rangle_{11\to 22}\sim 10^{-15}~{\rm cm^3/s}$ at Galactic
velocities.The approximation $q_\gamma\simeq 2R_\uparrow$ holds provided radiative decay dominates over collisional de-excitation,
$\Gamma_{\chi_2} \gg n_2 \langle\sigma v\rangle_{22\to 11}$ or equivalently $R_\downarrow\ll \Gamma_{\chi_2} n_2$;
we verify this condition for our benchmark throughout the Milky Way halo.

Dense molecular clouds (MCs) provide a sensitive probe of sub-keV photon
injection, since UV/EUV photons are efficiently absorbed in neutral gas and
produce primary and secondary ionizations. We compare our predicted EUV
injection against the MC ionization limits derived in
\cite{DelaTorreLuque:2025zjt}.
The ionization rate per ${\rm H}_2$ molecule at position $\mathbf{x}$ can be
written as 
\begin{equation}
\zeta_{{\rm H}_2}(\mathbf{x}) \;=\;
\int_{I_{{\rm H}_2}}^{\infty} dE\,
J_\chi(E,\mathbf{x})\,
\sigma^{\rm ion}_{{\rm H}_2}(E)\,
\Bigl[1+\theta_e(E)\Bigr],
\label{eq:zeta-def}
\end{equation}
where $I_{{\rm H}_2}=15.4~{\rm eV}$, $\sigma^{\rm ion}_{{\rm H}_2}(E)$ is the
photoionization cross section, and the number of secondary ionizations per
primary ionization is parameterized by
\begin{equation}
\theta_e(E)\;=\;\frac{E-I_{{\rm H}_2}}{W},
\qquad
W\simeq 40~{\rm eV},
\label{eq:thetae}
\end{equation}
with $W$ the mean energy loss per ion pair in the interstellar medium.

In our model the injected photon spectrum is monochromatic at
$E_\gamma\simeq\Delta m$, so the source term can be written as
$Q(E,\mathbf{x}) = q_\gamma(\mathbf{x})\,\delta(E-E_\gamma)$.
we convert the volume emissivity
into an in-cloud photon flux via the one-zone relation
\begin{equation}
J_\chi(E,\mathbf{x}) \;=\; 2\,Q(E,\mathbf{x})\,\frac{N_{{\rm H}_2}}{n_{{\rm H}_2}}
\;\simeq\; 2\,Q(E,\mathbf{x})\,L_{\rm MC},
\label{eq:Jchi-onezone}
\end{equation}
where $n_{{\rm H}_2}$ is the characteristic ${\rm H}_2$ density in the cloud,
$N_{{\rm H}_2}$ its column density, and $L_{\rm MC}\sim N_{{\rm H}_2}/n_{{\rm H}_2}$
its characteristic size.

Evaluating Eq.~\eqref{eq:zeta-def} for monochromatic injection yields
\begin{equation}
\zeta_{{\rm H}_2}(\mathbf{x})
\;\simeq\; J_\chi(E_\gamma,\mathbf{x})\,
\sigma^{\rm ion}_{{\rm H}_2}(E_\gamma)\,
\left(1+\frac{E_\gamma-I_{{\rm H}_2}}{W}\right),
\label{eq:zeta-mono}
\end{equation}
with $J_\chi(E_\gamma,\mathbf{x})\simeq 2q_\gamma(\mathbf{x})L_{\rm MC}$.
We apply this expression to the cloud targets and limits compiled in
\cite{DelaTorreLuque:2025zjt}, using the same assumptions for the DM density:
for local clouds we take the local DM density, while for inner-Galaxy clouds we
adopt an NFW profile as in \cite{DelaTorreLuque:2025zjt}. We find that our
predicted $\zeta_{{\rm H}_2}$ lies below the robust limits from well-characterized
clouds such as L1551 and the DRAGON cloud. We present results for the
inner-Galaxy cloud G1.4--1.8+87 separately, noting that it is treated as an
optimistic forecast target in \cite{DelaTorreLuque:2025zjt} due to
astrophysical systematics.
Related probes of soft photon injection at high redshift include 21-cm effects from decaying light DM producing Lyman-series photons \cite{Agius:2025nfz} ;
in our benchmark $\tau_{\chi_2}\sim 6\times10^3$~yr $\ll t(z\sim 20)$, so primordial $\chi_2$ decays are complete well before cosmic dawn, and any late-time signal is dominated by Milky-Way upscattering.
For keV--MeV splittings, X-ray searches provide complementary constraints on excited-state decays \cite{Krnjaic:2025zjl} , but these do not directly apply at $E_\gamma\sim 100$~eV.

\begin{figure}[htbp]
    \centering
    \includegraphics[width=0.95\linewidth]{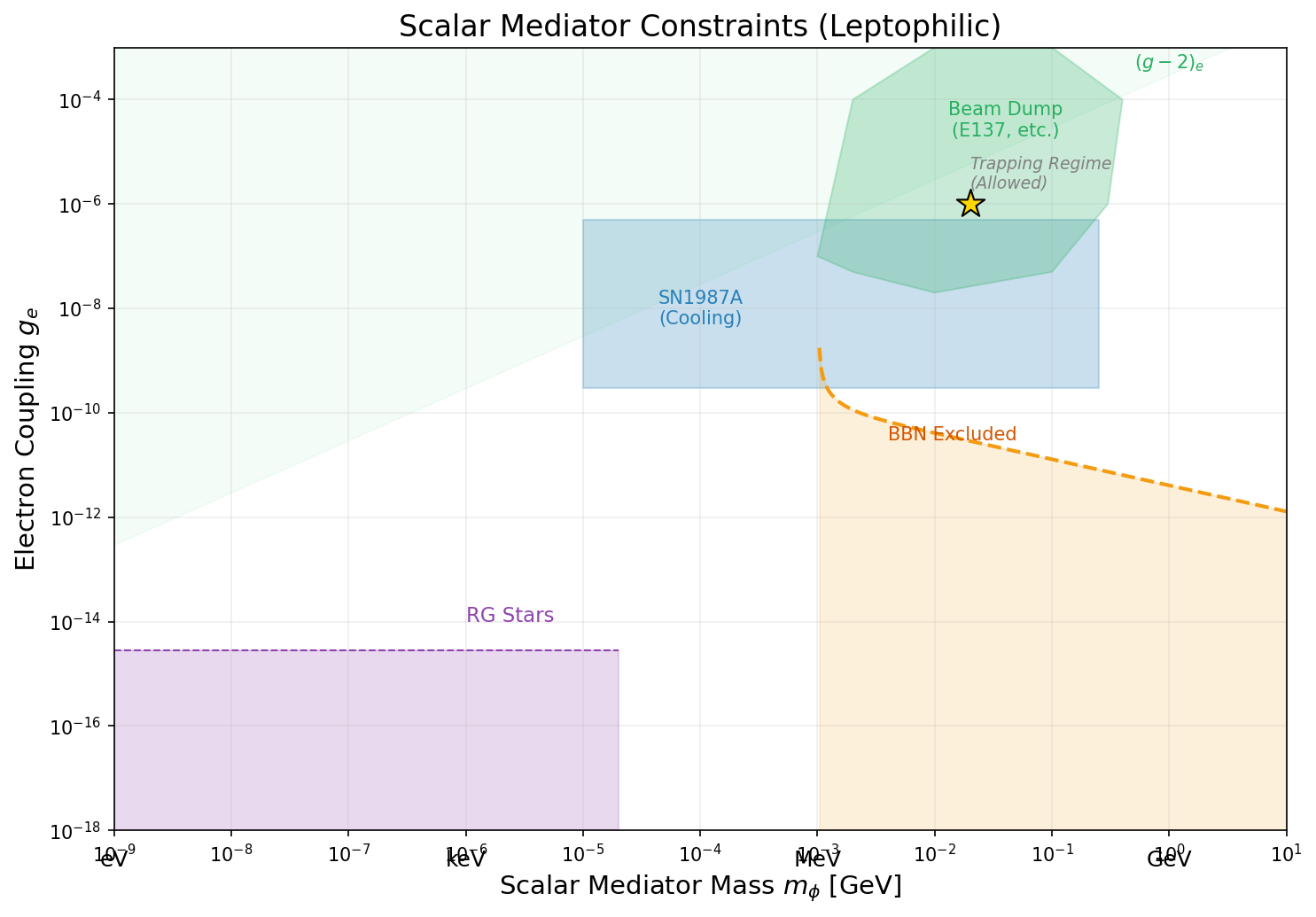}
    \caption{Combined cosmological and astrophysical constraints on the leptophilic scalar mediator $\phi$. The shaded regions indicate exclusion by: stellar cooling in Horizontal Branch/Red Giant stars (Purple)~\cite{Hardy:2016kme}; free-streaming cooling of SN1987A (Blue)~\cite{Chang:2018rso}; visible decay searches in beam-dump experiments (Green)~\cite{Liu:2016qwd}; and Big Bang Nucleosynthesis (BBN) disruption for lifetimes $\tau_\phi > 1$ s (Orange)~\cite{Cyburt:2015mya}. The gold star marks our benchmark ($m_\phi = 20$ MeV, $g_e = 10^{-6}$), which resides in the SN1987A trapping regime where scalars are trapped in the core and decays promptly ($\tau \ll 1$ s) to ensure BBN safety, while remaining below the sensitivity of past beam-dump searches.}. 
    \label{fig:cosmological_constraints}
\end{figure}
\begin{figure}[htbp]
    \centering
    \includegraphics[width=0.95\linewidth]{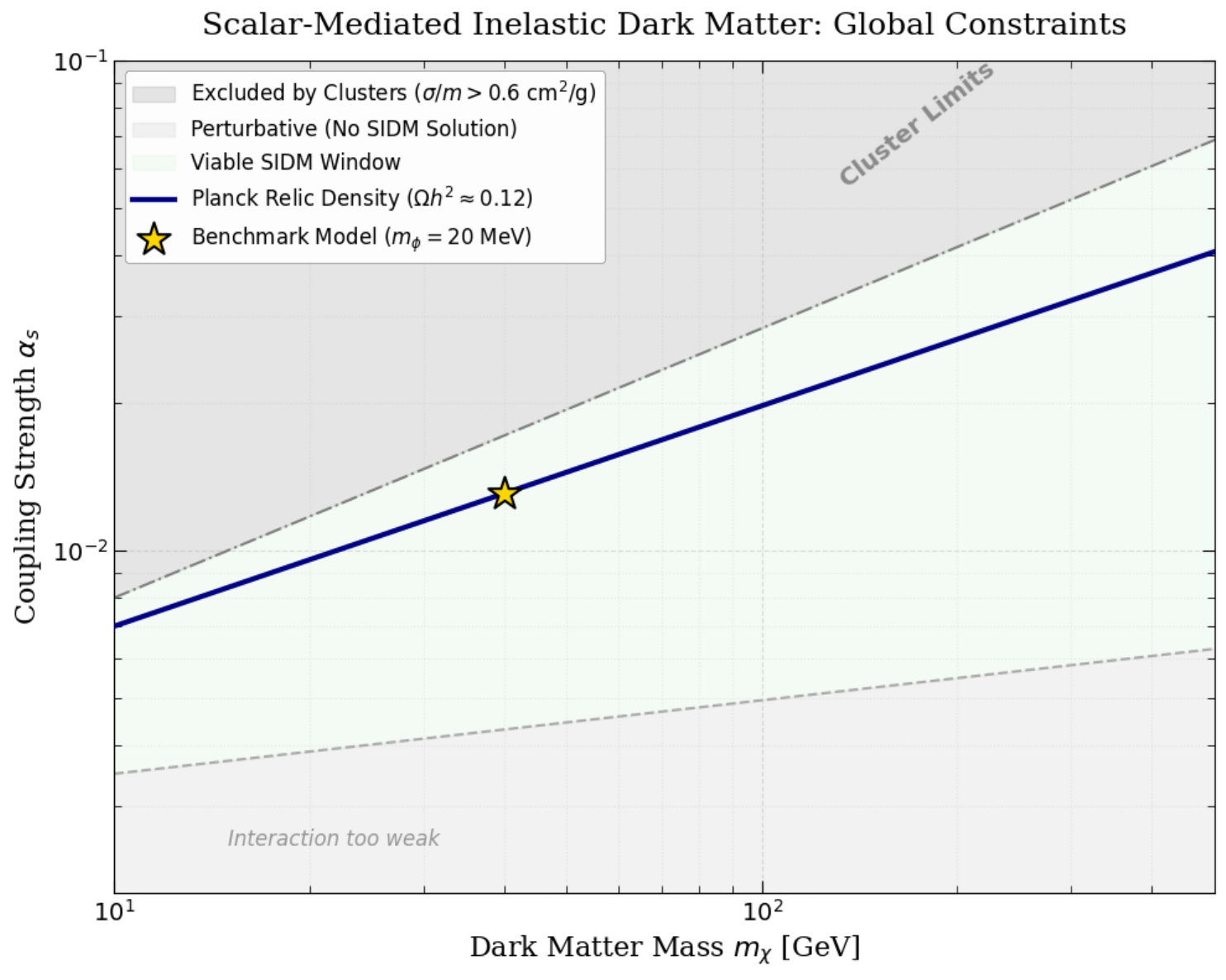}
    \caption{The discovery window for $M_{\chi}$and $\alpha$, combining bounds from relic density\cite{Planck2018} , non-perturbative limits,below it the non-perturbative treatment failed, and cluster exclusion\cite{Randall2008}}
    \label{}
\end{figure}
\section{Benchmark and Allowed Parameter Space}
\label{sec:results}

We performed a comprehensive Monte Carlo parameter scan using the full non-perturbative coupled-channel calculation (see Appendix~\ref{app:scattering}). We imposed conditions for satellite suppression (Draco, $\sigma < 0.2$ cm$^2$/g), dwarf galaxy solution ($\sigma \sim 1-10$ cm$^2$/g), and cluster constraints ($\sigma < 0.6$ cm$^2$/g).

At cluster velocities ($v \sim 10^3$ km/s), the non-perturbative calculation yields a transport cross-section $\sigma_T/m \approx 0.05$ cm$^2$/g. This value is safely below the robust upper bound of $\sigma/m < 0.6$ cm$^2$/g derived from cluster halo ellipticities and merging clusters.

Our non-perturbative analysis identifies a resonant benchmark around
\begin{equation}
    \begin{split}
        m_\chi &\approx 40~\text{GeV}, \qquad m_\phi \approx 20~\text{MeV}, \\
        \quad \Delta m &\approx 100~\text{eV}, \qquad \alpha \approx 10^{-2}.
    \end{split}
\end{equation}
This point passes all astrophysical constraints via the resonance mechanism described in Sec.~\ref{sec:scattering}. 

\begin{figure}[htbp]
    \centering
    \includegraphics[width=0.95\linewidth]{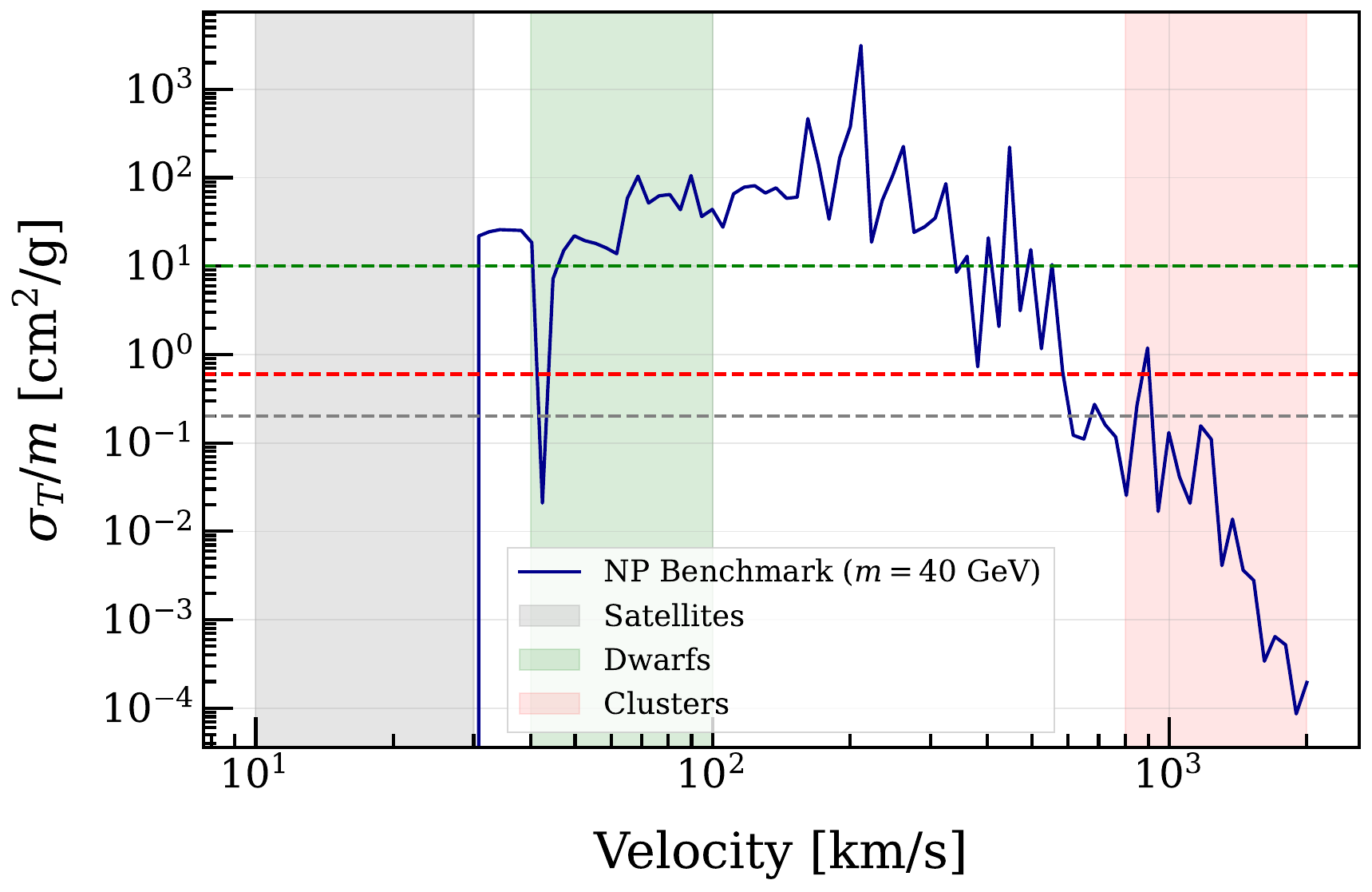}
    \caption{The velocity-dependent transport cross-section $\sigma_T/m$ for the scalar inelastic model (Benchmark: $m_\chi=40$ GeV, $\Delta m=100$ eV). The calculation uses the full non-perturbative Schrödinger solution. The cross-section is suppressed in the satellite regime (gray), resonates in the dwarf regime (green), and is suppressed in the cluster regime (red)}
    \label{fig:sigma_v}
\end{figure}

\begin{figure}[htbp]
    \centering
    \includegraphics[width=0.95\linewidth]{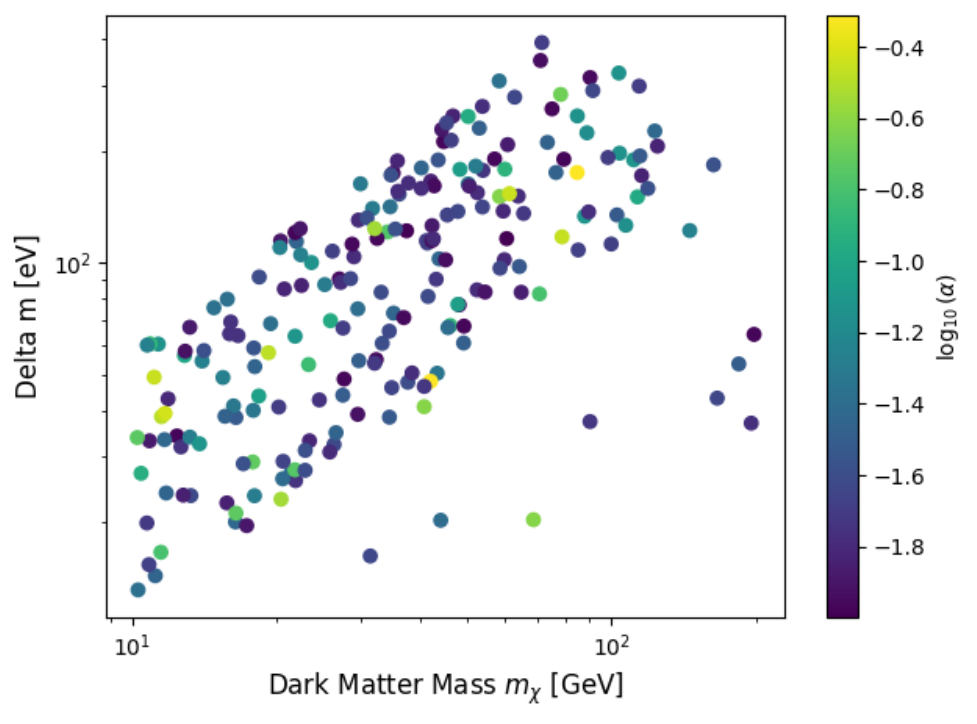}
    \caption{Allowed parameter space from non-perturbative analysis. The color scale indicates the coupling strength $\log_{10}(\alpha)$. The points represent models that satisfy all self-interaction constraints. }
    \label{fig:param_space}
\end{figure}

Our scan identifies a distinct island of viability in the $m_\chi \text{--} \Delta m$ plane. Fig.~\ref{fig:param_space} illustrates the allowed region where $\Delta m$ scales with $m_\chi$ to maintain the kinematic threshold.

\section{Direct Detection Phenomenology}
\label{sec:dd_detailed}

The direct detection phenomenology of our model is different from standard Higgs-portal scenarios because its the scalar mediator is leptophilic . The leptophilic portal induces an effective dark matter charge radius and anapole moment at the loop level  via $\chi$-$\chi$-$\phi$-$\ell$-$\ell$ diagrams~\cite{Kopp:2014tsa}. Dimensional analysis suggests these contributions scale as $g_S g_\ell e/({16\pi^2 m_\phi^2})$. Compared to the transition dipole operator $1/\Lambda$ (where $\Lambda_eff \sim 10^7$ GeV), these loop effects are suppressed by factors of the lepton mass and loop factors,thus they are sub-dominant to the tree-level dipole scattering.

\subsection{Standard electronic and nuclear recoils}

In standard scalar SIDM models, the mixing with the Higgs boson generates a spin-independent cross-section on nucleons, $\sigma_\text{SI}$. For a mixing angle $\sin\theta \sim 10^{-4}$ required for BBN in the standard scenario, this cross-section is typically $\sim 10^{-36}$ cm$^2$, which is firmly excluded by underground direct search experiments for $m_\chi \approx 40$ GeV.The $\Delta m \sim 100$ eV splitting is also insufficient to kinematically block the scattering~\cite{LZ:2022lsv, XENON:2025vwd, PandaX:2024qfu}. However,in leptophilic scenario the tree-level coupling to quarks is absent ($g_q = 0$). Scattering off nuclei can only occur via loop processes like a photon loop involving the scalar-electron loop.These processes are loop-suppressed and negligible, or via the magnetic dipole operator. 
Inelastic up-scattering $\chi_{1}N\rightarrow\chi_{2}N$ is kinematically accessible in terrestrial detectors. However, to satisfy current constraints from liquid xenon experiments, the dipole suppression scale must be $\Lambda_{eff} \ge 10^7$ GeV. This implies a macroscopic lifetime for the excited state:
\begin{equation}
\tau_{\chi_2} \simeq \frac{\pi \Lambda_{eff}^2}{\Delta m^3} \sim 10^{11} \text{ s}.
\end{equation}
Even at characteristic dark matter velocities ($v \sim 10^{-3}c$), the decay length is astronomical ($L \gg 10^8$ km).  The excited state escapes the detector volume long before decaying. The "luminous" photon signature is therefore not observable in standard setups.

Instead, the primary discovery mode is the unique nuclear recoil spectrum induced by the massless photon mediator. Unlike contact interactions which yield a flat recoil spectrum at low energies, the dipole operator induces a differential cross-section scaling roughly as $E_R^{-1}$:
\begin{equation}
\frac{d\sigma}{dE_{R}} \approx \frac{4\pi\alpha_{EM}Z^{2}}{\Lambda_{eff}^{2}} \frac{1}{E_{R}} \left(1 - \frac{v_{min}^2}{v^2}\right) F_{Helm}^2(E_R).
\end{equation}
This results in a sharp rise in the event rate at low recoil energies, strictly truncated by the inelastic kinematic threshold. Future low-threshold experiments (e.g., SuperCDMS) will search for this characteristic spectral shape rather than a mono-chromatic photon.
Furthermore, even if the leptophilic portal allows for $\chi e \rightarrow \chi e$ scattering, with the corrected coupling $g_e \approx 10^{-6}$, the cross-section is highly suppressed. Therefore, electronic recoil (ER) rate of our model is negligible compared to current limits in Ref.~\cite{XENON:2022ltv, LZ:2023poo}, leaving the magnetic dipole nuclear recoil channel as the primary discovery mode. 

\subsection{Magnetic dipole nuclear recoil}

From previous discussion, the primary direct detection signature arises from the transition magnetic dipole operator. This induces inelastic scattering $\chi_1 N \rightarrow \chi_2 N$ mediated by photon exchange. The differential cross-section is
\begin{equation}
    \begin{split}
        \frac{\d\sigma}{\d E_R} &= \frac{4\pi \alpha_\text{EM} Z^2}{\Lambda_{eff}^2} \frac{1}{E_R} \times \\
        &\qquad \left[ 1 - \frac{E_R}{v^2} \frac{m_N + 2m_\chi}{2m_\chi m_N} - \frac{\Delta m}{v^2 \mu_{\chi N}} \right] F^2_\text{Helm}(E_R),
    \end{split}
\end{equation}
which gives a detection rate
\begin{equation}
    \frac{\d R}{\d E_R} = N_T\,\frac{\rho_{\chi}}{m_\chi} \int_{v_\text{min}(E_R)}^\infty \frac{\d\sigma}{\d E_R}\, v f(v)\ \d v.
\end{equation}

For our benchmark $\Delta m = 100$ eV, the minimum velocity $v_\text{min}$ required to produce a recoil $E_R$ is significantly increased compared to elastic scattering:
\begin{equation}
    v_\text{min} = \sqrt{\frac{1}{2m_N E_R}} \left( \frac{m_N E_R}{\mu_{\chi N}} + \Delta m \right),
\end{equation}
For a given $\Lambda_{\rm eff}$, the excited-state lifetime is $\tau_{\chi_2}\sim \pi\,\Lambda_{\rm eff}^2/\Delta m^3$,
so for sufficiently large $\Lambda_{\rm eff}$ the excited states produced in the detector escape before decaying
($c\tau \gg L_{\rm det}$).  
The scattering cross-section via the dipole operator includes two terms: the coherent Dipole-Charge interaction ($\propto Z^2$) and the spin-dependent Dipole-Dipole interaction ($\propto \mu_N^2$).
\begin{equation}
\frac{d\sigma}{dE_R} \propto \frac{1}{E_R} \left[ Z^2 F_{charge}^2(E_R) + \frac{J+1}{3J} \mu_N^2 F_{spin}^2(E_R) \right].
\end{equation}
For Xenon ($Z=54$), the coherent charge term dominates by a factor of $\sim 10^4$. We utilize this enhanced response in our projections. However, the stability requirement $\Lambda_{eff} \ge 10^7$ GeV suppresses the total rate to $< 10^{-6}$ events/ton-year, rendering it unobservable in current facilities.
\subsection{Dipole Nuclear Response and Experimental Reach}

Following the formalism of Magnetic Inelastic Dark Matter (MiDM) \cite{Chang:2010en,Fitzpatrick:2012ix}, the differential cross-section for dipole-induced scattering is the sum of a Dipole-Charge (DZ) term and a Dipole-Dipole (DD) term:
\begin{equation}
    \frac{d\sigma}{dE_R} = \frac{d\sigma_{\text{DZ}}}{dE_R} + \frac{d\sigma_{\text{DD}}}{dE_R}.
\end{equation}
For heavy target nuclei like Xenon ($Z=54$), the Charge term dominates due to the coherent $Z^2$ enhancement:
\begin{equation}
\begin{split}
    \frac{d\sigma_{\text{DZ}}}{dE_R} \approx \frac{4\pi \alpha_{\text{EM}} Z^2}{E_R \Lambda_{\text{eff}}^2}\\ \qquad\left[ 1 - \frac{E_R}{v^2} \left( \frac{m_N + 2m_\chi}{2m_\chi m_N} \right) - \frac{\delta}{v^2 \mu_{\chi N}} \right] |F_{\text{ch}}(E_R)|^2,
    \end{split}
\end{equation}
where $F_{\text{ch}}(E_R)$ is the nuclear charge form factor. The Dipole-Dipole term, proportional to the nuclear magnetic moment $\mu_N^2$, is subdominant for spin-suppressed isotopes (e.g., $^{132}$Xe) and is neglected in our projection, providing a conservative estimate of the rate.

\subsection{Results and Future Detectability}
Current direct detection experiments such as XENON1T have set stringent limits on the dark matter magnetic dipole moment, excluding $\mu_\chi \gtrsim 10^{-5} \mu_N$ for mass splittings $\Delta m \lesssim 100$ keV. Our benchmark scale $\Lambda_{\text{eff}} \sim 10^7$ GeV corresponds to an effective moment $\mu_\chi \approx e/\Lambda_{\text{eff}} \sim 2 \times 10^{-7} \mu_N$. Since the event rate scales as $\mu_\chi^2 \propto \Lambda_{\text{eff}}^{-2}$, our benchmark signal is suppressed by a factor of $\sim 2500$ relative to the current exclusion threshold. This confirms that the model lies deep within the allowed region.

To substantiate the discovery potential, we project the event rates for next-generation experiments by integrating the differential cross-section over the detector efficiency $\epsilon(E_R)$. We assume a sharp threshold efficiency to illustrate the scaling.

The total event rate scales with the dipole suppression scale as $R \propto \Lambda_{eff}^{-2}$. Calibrating to the current XENON1T exclusion limit $\Lambda_{eff} \gtrsim 2 \times 10^5$ GeV, corresponding to $\mathcal{O}(10)$ events/ton-year, we obtain the following projection for a Liquid Xenon target (LZ/XLZD) with $E_{th} = 1$ keV$_{nr}$:
\begin{equation}
    R_{Xe} \approx 0.4 \left( \frac{10^6 \text{ GeV}}{\Lambda_{eff}} \right)^2 \text{ events/(ton}\cdot\text{year)}.
    \label{eq:rate_proj}
\end{equation}
For our benchmark ($\Lambda_{eff} = 10^7$ GeV), the rate is $\sim 4 \times 10^{-3}$ events/(ton$\cdot$year), which is effectively unobservable. However, a Discovery Benchmark of $\Lambda_{eff} = 10^6$ GeV remains consistent with all current null results  but yields $\sim 40$ signal events in a 100 ton-year exposure of XLZD. This signal magnitude would be statistically significant above the coherent neutrino scattering floor.

For low-threshold detectors like SuperCDMS (Germanium, $Z=32$), the lack of $Z^2$ enhancement is compensated by the ability to probe the $1/E_R$ divergence. Assuming $E_{th} = 100$ eV$_{nr}$, we project:
\begin{equation}
    R_{Ge} \approx 0.2 \left( \frac{10^6 \text{ GeV}}{\Lambda_{eff}} \right)^2 \text{ events/(ton}\cdot\text{year)}.
\end{equation}
While the integral rate is slightly lower than Xenon, the observation of the spectral rise below 1 keV would constitute a smoking-gun signature distinguishing the dipole operator from contact interactions or neutrino backgrounds. Thus, the model offers a clear target for the multi-ton scale era, provided the dipole scale lies in the $\Lambda_{eff} \sim 10^6$ GeV window.

The integrated event rates per unit mass are comparable between the two technologies. The overall normalization requires $\Lambda_{\text{eff}}$ in the multi-PeV range to satisfy null results, leading to rates of $\sim 10^{-7}$ events/(ton$\cdot$yr) in xenon and $\sim 10^{-6}$ events/(ton$\cdot$yr) in germanium for our benchmark. While these rates are unobservable at present exposures, future low-threshold searches can improve sensitivity by probing the characteristic $1/E_R$ spectral shape. A signal in SuperCDMS exhibiting the $E_R^{-1}$ rise, combined with a kinematically suppressed signal in LZ, would provide a distinct fingerprint for this dipole inelastic scenario.

\begin{figure}[h]
    \centering
    \includegraphics[width=0.95\linewidth]{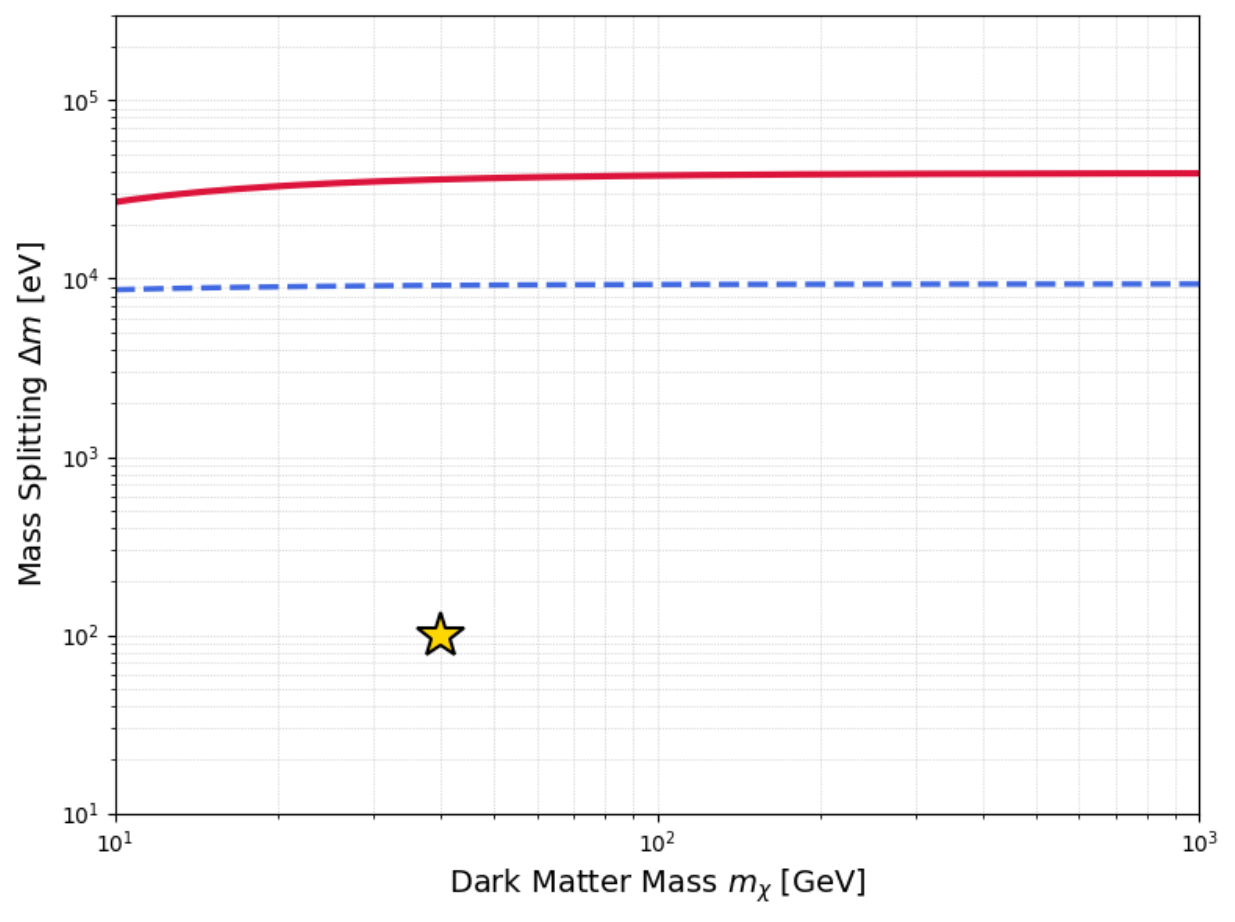}
    \caption{Kinematic accessibility for inelastic dipole dark matter.
The solid red line is the maximum mass splitting $\Delta m$ accessible to liquid xenon experiments
LZ, PandaX-4T for a representative nuclear-recoil threshold .
The dashed blue line shows the corresponding reach for germanium-based detectors  SuperCDMS
assuming $E_{\rm th}=0.1~\mathrm{keV_{nr}}$ and the lighter target nucleus ($m_{\rm Ge}\simeq 72.6~\mathrm{GeV}$).
The star marks our resonant benchmark , which is
kinematically accessible to both targets. In the direct-detection-safe regime the overall rate
is nevertheless strongly suppressed by the large dipole scale, scaling as $R\propto \Lambda_{\rm eff}^{-2}$;
lowering $E_{\rm th}$ primarily enlarges the accessible $\Delta m$ range and improves sensitivity to the
characteristic $1/E_R$ recoil spectrum.}
    \label{fig:dd_prospects}
\end{figure}

In Figure~\ref{fig:dd_prospects}, we illustrate the kinematic accessibility of the inelastic channel in the $(m_\chi,\Delta m)$ plane. The solid red curve denotes the maximum splitting accessible to liquid xenon ($E_{\text{th}}=1$ keV$_{\text{nr}}$), while the dashed blue curve shows the reach for germanium ($E_{\text{th}}=0.1$ keV$_{\text{nr}}$). Our resonant benchmark $(m_\chi,\Delta m)=(40~\text{GeV},\,100~\text{eV})$ lies well below both curves, confirming that inelastic nuclear recoils are kinematically permitted in both detector types, limited only by the interaction strength $\Lambda_{\text{eff}}$.

\section{Robustness of the Benchmark Model}

We analyze the stability of our chosen benchmark point ($m_\chi = 40$ GeV, $m_\phi = 20$ MeV, $\alpha \approx 0.013$) against the global constraints. The benchmark lies within the viable SIDM window, satisfying all cosmological and astrophysical bounds. Below, we quantify the safety margins for the key model parameters, defining the percentage variation allowed before the model encounters an exclusion limit.

\subsection{Coupling Strength ($\alpha$)}
The dark sector coupling strength $\alpha$ is the most constrained parameter, bounded from above by cluster halo ellipticity\cite{Bautista:2025lxk} constraints and from below by the requirement to solve small-scale structure problems.
The upper bound is contributed by cluster limits .The benchmark sits at $\alpha \approx 0.013$. The exclusion limit from cluster scales ($\sigma/m > 0.6$ cm$^2$/g) at $m_\chi = 40$ GeV lies at approximately $\alpha^{\text{max}} \approx 0.017$. This provides a safety margin of:
    \begin{equation}
        \frac{\alpha^{\text{max}} - \alpha}{\alpha} \approx \frac{0.017 - 0.013}{0.013} \approx +30\%.
    \end{equation}
    Increasing the coupling beyond this value would result in excessive scattering at high velocities ($v \sim 1000$ km/s), potentially disrupting galaxy cluster cores.The minimum coupling required to address the core-cusp problem in dwarf galaxies is approximately $\alpha^{\text{min}} \approx 0.004$. 
    \begin{equation}
        \frac{\alpha - \alpha^{\text{min}}}{\alpha} \approx \frac{0.013 - 0.004}{0.013} \approx -70\%.
    \end{equation}
    This indicates the model remains viable even if the coupling is reduced by nearly a factor of 3.

\subsection{Dark Matter Mass ($m_\chi$) and Mediator Mass ($m_\phi$)}
The mass parameters have a high degree of robustness.The viable window extends from $m_\chi \sim 5$ GeV to well above $1$ TeV. Our benchmark at 40 GeV is not fine-tuned; Largely shifting $m_\chi$  keeps the model within the safe region according to the combined constraint diagram.For the mediator mass $m_\phi = 20$ MeV is effectively bounded by the BBN constraint on the lower end. To ensure decay into SM particles well before nucleosynthesis, we require $m_\phi \gg 2m_e \approx 1$ MeV. To avoid interfering the neutrino decoupling we need a safe floor of $m_\phi \sim 10$ MeV. Thus, the parameter space allows a variation of roughly $-50\%$ and $+400\%$ up to the $\sim 100$ MeV scale where the secluded approximation begins to fail.
As discussed in Sec\ref{sec:cosmo}, it also naturally evades CMB bounds through S-wave annihilation. Therefore,we can sure that its is a robust SIDM model that do not rely on careful fine-tuning to pass constraints.

\section{Discussion and Conclusion}
\label{sec:conclusion}

In this work, we have proposed a scalar-mediated inelastic Dark Matter model that provides a robust and self-consistent solution to the small-scale structure anomalies of the $\Lambda$CDM paradigm. By enforcing a discrete $\mathbb{Z}_2$ symmetry and solving the non-perturbative Schrödinger dynamics, we identified  resonant benchmark around ($m_\chi \approx 40$ GeV, $\Delta m \approx 100$ eV) that successfully reconciles the cored density profiles of dwarf galaxies with the survival of dense satellite halos. This phenomenology relies on the velocity-dependent saturation of the Sommerfeld enhancement, which naturally suppresses interactions in the ultra-faint satellite regime while resonating at the velocity scales characteristic of dwarf galaxies.

Our framework is distinguished from existing literature for  three phenomenological advantages:
\begin{itemize}
    \item[(i)] Cosmological consistency via $p$-wave annihilation: Unlike vector-mediated SIDM constrained by $s$-wave annihilation limits from the CMB~\cite{Tulin2013}, our scalar mediator naturally exhibits $p$-wave suppression. This feature allows the model to accommodate the large self-interaction cross-sections required for halo cores without conflicting with Planck data or requiring fine-tuning. Additionally, unlike freeze-in scenarios characterized by metastable excited states\cite{Krnjaic:2025zjl}, our freeze-out framework
does not rely on a rapid de-excitation of $\chi_2$.
In the direct-detection-safe regime we typically require $\Lambda_{\rm eff}\gtrsim 10^{7}~\mathrm{GeV}$,
for which $\chi_2$ can be long-lived. This is nevertheless cosmologically safe because the injected
electromagnetic energy per decay is only $\Delta m\ll \mathrm{MeV}$, far below photodissociation thresholds,
and the fractional energy release satisfies $\Delta m/m_\chi\ll 1$.
    \item[(ii)] Robust kinematic switch via hard symmetry: A critical feature of our model is the imposition of a hard $\mathbb{Z}_2$ symmetry to forbid tree-level elastic scattering, rather than relying on spontaneous symmetry breaking or effective field theories~\cite{Bringmann:2016din, Ref2024}. This ensures that the kinematic suppression in satellites is absolute, protecting ultra-faint halos from the runaway gravothermal collapse known as Nishikawa instability that affects models with residual elastic leakage.
    \item[(iii)] The dipole divergence spectral signature: A salient feature of this scenario is its unique direct detection phenomenology. While standard inelastic models mediated by massive vector bosons predict contact-like interactions with flat recoil spectra, our dimension-5 transition magnetic dipole operator induces scattering mediated effectively by the photon. This results in a differential cross-section that scales as $\d\sigma/\d E_R \propto E_R^{-1}$. This dipole divergence manifests as a sharp enhancement in the event rate at low recoil energies, strictly truncated by the inelastic kinematic threshold.
\end{itemize}

In the context of current experimental constraints, the model remains viable. While current xenon-based experiments have placed stringent limits on dark matter-nucleus scattering~\cite{LZ:2022lsv, XENON:2025vwd, PandaX:2024qfu}, our model evades these bounds due to the leptophilic nature of the scalar mediator which eliminates tree-level quark couplings and the suppression of rates in xenon targets due to the inelastic threshold. The primary discovery potential lies with next-generation xenon-based experiments such as XLZD~\cite{XLZD:2024nsu}. The observation of a recoil spectrum characterized by the distinct $E_R^{-1}$ enhancement would constitute smoking-gun evidence for this scenario, distinguishing it from standard contact-interaction dark matter. \newline
Finally, we comment on the genesis of this dark sector. The mass hierarchy required for the SIDM solution ($m_\chi \gg m_\phi$) suggests a dynamical origin involving a phase transition in the early universe. Preliminary considerations suggest such a transition could occur near the BBN scale. It could potentially generate a stochastic gravitational wave background in the nanohertz frequency range. We reserve a detailed study of this cosmological history and its connection to Pulsar Timing Array signals for future work.

\appendix

\section{Formalism for Non-Perturbative Inelastic Scattering}
\label{app:scattering}

In this section, we present the detailed calculation of the scattering cross-section in the non-perturbative regime. For the process $\chi_1 \chi_1 \rightarrow \chi_2 \chi_2$ mediated by a scalar $\phi$, the potential matrix in the basis $(\chi_1 \chi_1, \chi_2 \chi_2)$ is off-diagonal.

\subsection{Coupled-channel Schrödinger equation}

Consider the two-body wavefunction $\Psi(\vec{r}) = (\psi_1(\vec{r}), \psi_2(\vec{r}))^T$, where channel 1 is $\chi_1\chi_1$ and channel 2 is $\chi_2\chi_2$. The radial Hamiltonian for a partial wave $l$ is:
\begin{equation}
    \left[ \frac{\d^2}{\d r^2} + K^2 - \frac{l(l+1)}{r^2} - 2\mu \boldsymbol{V}(r) \right] \boldsymbol{u}_l(r) = 0,
\end{equation}
where $\boldsymbol{u}_l = (u_{l,1}, u_{l,2})^T$. The momentum matrix $K^2$ is diagonal:
\begin{equation}
    K^2 = \begin{pmatrix} k_1^2 & 0 \\ 0 & k_2^2 \end{pmatrix},
\end{equation}
with $k_1^2 = m_\chi E$ and $k_2^2 = m_\chi (E - 2\Delta m)$. The interaction potential matrix induced by Eq.~\eqref{eq:scalar_lagrangian} is Hermitian and strictly off-diagonal:
\begin{equation}
    \boldsymbol{V}(r) = \begin{pmatrix} 0 & -\frac{\alpha}{r}\mathrm e^{-m_\phi r} \\ -\frac{\alpha}{r} \mathrm e^{-m_\phi r} & 0 \end{pmatrix}.
\end{equation}
Note that $V_{12} = V_{21}$, ensuring the Hamiltonian is Hermitian. The minus sign arises from the attractive nature of the scalar exchange in the $t$-channel for the $\chi_1 \chi_1 \rightarrow \chi_2 \chi_2$ transition.

To obtain numerical solution, we transform to dimensionless variables. We define $x = \alpha m_\chi r$ and the dimensionless parameters:
\begin{equation}
    \epsilon_v = \frac{v_\text{rel}}{2\alpha}, \quad \epsilon_\phi = \frac{m_\phi}{\alpha m_\chi}, \quad \epsilon_\delta = \sqrt{\frac{4\Delta m}{m_\chi \alpha^2}}.
\end{equation}
The coupled radial equations become:
\begin{align}
    \left[ \frac{\d^2}{\d x^2} + \epsilon_v^2 - \frac{l(l+1)}{x^2} \right] u_{l,1} &= -\frac{2}{x}\mathrm e^{-\epsilon_\phi x} u_{l,2}, \\
    \left[ \frac{\d^2}{\d x^2} + (\epsilon_v^2 - \epsilon_\delta^2) - \frac{l(l+1)}{x^2} \right] u_{l,2} &= -\frac{2}{x}\mathrm e^{-\epsilon_\phi x} u_{l,1},
\end{align}
which can be solved numerically. 

\subsection{Boundary conditions and S-matrix}

We solve these equations for the S-matrix element $S_{12}$ describing the transition $1 \rightarrow 2$.
\begin{itemize}
    \item \emph{At the origin ($x \rightarrow 0$):} The wavefunction must be regular, $u_{l,i} \sim x^{l+1}$.
    \item \emph{At infinity ($x \rightarrow \infty$):} The potential vanishes. For an incoming wave in channel 1, we match to the asymptotic form:
    \begin{align}
        u_{l,1} &\sim \frac{\mathrm i}{2} \left( h_l^{(2)}(k_1 r) - S_{11} h_l^{(1)}(k_1 r) \right), \\
        u_{l,2} &\sim -\frac{\mathrm i}{2} \sqrt{\frac{k_1}{k_2}} S_{12} h_l^{(1)}(k_2 r),
    \end{align}
    where $h_l^{(1,2)}$ are spherical Hankel functions.
\end{itemize}
Below threshold ($k_2^2 < 0$), $k_2$ becomes imaginary, $k_2 \rightarrow\mathrm i \kappa_2$. The wavefunction in channel 2 must decay exponentially, $u_{l,2} \sim \mathrm e^{-\kappa_2 r}$, implying $S_{12} = 0$.

To verify the numerical stability of the coupled-channel solver, we explicitly check the unitarity of the S-matrix. We define the flux in channel $i$ as $\mathcal{F}_i \propto k_i |S_{1i}|^2$. Above the inelastic threshold, flux conservation requires $\sum_i k_i |S_{1i}|^2 = k_1$. Our numerical solutions satisfy this condition to within $10^{-5}$ precision, confirming that the effective non-Hermitian appearance (due to open channels) is correctly handled by the complex S-matrix matching procedure.

\subsection{Cross-section calculation}

For $i\rightarrow j$ scattering, the partial-wave amplitude is
\begin{equation}
    f_{ij}(\theta)=\frac{1}{2\mathrm i k_i}\sum_{l=0}^\infty (2l+1)\left(S_{ij}^{(l)}-\delta_{ij}\right)P_l(\cos\theta),
\end{equation}
and the differential cross section is
\begin{equation}
    \frac{\d\sigma_{i\rightarrow j}}{\d\Omega}=\frac{k_j}{k_i}\,|f_{ij}(\theta)|^2,
\end{equation}
with $k_j$ understood as real above threshold and $\d\sigma_{i\rightarrow j}=0$ below threshold.

We use the standard definitions
\begin{align}
    \sigma_T^{i\rightarrow j} &= \int \d\Omega\,(1-\cos\theta)\,\frac{\d\sigma_{i\rightarrow j}}{\d\Omega},\\
    \sigma_V^{i\rightarrow j} &= \int \d\Omega\,\sin^2\theta\,\frac{\d\sigma_{i\rightarrow j}}{\d\Omega},
\end{align}
and evaluate these integrals numerically using the $S$-matrix output. For identical particles in an elastic channel, the usual (anti-)symmetrization of $f(\theta)$ is applied. The inelastic scattering cross-section is obtained by summing the partial wave contributions:
\begin{equation}
    \sigma_{1\rightarrow 2} = \frac{\pi}{k_1^2} \sum_{l=0}^{\infty} (2l+1) |S_{12}^{(l)}|^2.
\end{equation}
For thermalization, we are interested in the viscosity cross-section $\sigma_T$. Since the final state particles are distinct from the initial state (and heavier), we define the energy transfer cross-section explicitly or use the standard weighted sum:
\begin{equation}
    \sigma_T \approx \frac{\pi}{k_1^2} \sum_{l=0}^{\infty} \frac{(l+1)(2l+1)}{2l+3} |S_{12}^{(l)}|^2.
\end{equation}
Numerical evaluation of this sum reveals the resonant structure discussed in Sec.~\ref{sec:scattering}.

\subsection{Numerical implementation and validation}
\label{app:numerics}

We then specify the numerical procedure used to obtain the
non-perturbative inelastic scattering cross sections shown in the main text.

%\subsubsection{Kinematics and channel momenta}
We solve the coupled-channel Schr\"{o}dinger equation in the basis $\left(\chi_1\chi_1,\chi_2\chi_2\right)$ with reduced mass $\mu \simeq {m_\chi}/2$. For a given relative velocity $v_{\rm rel}$ (in the center-of-mass frame), the kinetic energy in the incoming channel is approximately $E \equiv \frac{\mu v_{\rm rel}^2}{2}$. The outgoing inelastic channel is heavier by $2\Delta m$, hence its available kinetic energy is $E-2\Delta m$. The channel momenta are
\begin{equation}
    \begin{split}
        k_1 &= \sqrt{2\mu E} = \mu v_{\rm rel},\\
        k_2 &=
        \begin{cases}
            \sqrt{2\mu (E-2\Delta m)}, & (E>2\Delta m),\\
        \mathrm i\,\kappa_2,\quad \kappa_2\equiv \sqrt{2\mu(2\Delta m-E)}, & (E<2\Delta m).
        \end{cases}
    \end{split}
\end{equation}
The inelastic channel opens only above the threshold velocity
\begin{equation}
    v_{\rm th}=\sqrt{\frac{8\Delta m}{m_\chi}}.
\end{equation}
Below threshold we enforce an exponentially decaying solution in channel 2 and the physical inelastic transition probability vanishes, $S_{12}^{(l)}=0$.

%\subsubsection{Radial ODE system and initial conditions}
For each partial wave $l$ we write the coupled radial equations as a first-order system for the vector
\begin{equation}
    y(r) \equiv \left(u_1(r),u_1'(r),u_2(r),u_2'(r)\right)^T .
\end{equation}
Using the potential matrix in Eq.~(A3),
\begin{equation}
    \boldsymbol{V}(r)=
    \begin{pmatrix}
    0 & -\alpha \,\mathrm e^{-m_\phi r}/r\\
    -\alpha \,\mathrm e^{-m_\phi r}/r & 0
    \end{pmatrix},
\end{equation}
the system takes the form
\begin{align}
    u_1''(r) &= \left[\frac{l(l+1)}{r^2}-k_1^2\right]u_1(r) + 2\mu V_{12}(r)\,u_2(r),\label{eq:ode_u1}\\
    u_2''(r) &= \left[\frac{l(l+1)}{r^2}-k_2^2\right]u_2(r) + 2\mu V_{21}(r)\,u_1(r).\label{eq:ode_u2}
\end{align}
We start the integration at a small but finite radius $r_{\min}$ and impose regular boundary conditions $u_{i}(r)\propto r^{l+1}$. Numerically we propagate two linearly independent solutions $(A,B)$:
\begin{align}
    y_A(r_{\min}) &= \left(r_{\min}^{l+1},\ (l+1)r_{\min}^{l},\ 0,\ 0\right)^T,\label{eq:icA}\\
    y_B(r_{\min}) &= \left(0,\ 0,\ r_{\min}^{l+1},\ (l+1)r_{\min}^{l}\right)^T.\label{eq:icB}
\end{align}
From these, at the matching point $r_{\max}$ we form the $2\times 2$ fundamental matrix of solutions and its derivative,
\begin{equation}
    \boldsymbol{M}(r_{\max})=
    \begin{pmatrix}
    u_{1,A} & u_{1,B}\\
    u_{2,A} & u_{2,B}
    \end{pmatrix},\quad
    \boldsymbol{M}'(r_{\max})=
    \begin{pmatrix}
    u_{1,A}' & u_{1,B}'\\
    u_{2,A}' & u_{2,B}'
    \end{pmatrix}.
\end{equation}

%\subsubsection{Integration method and domain}
We integrate Eqs.~\eqref{eq:ode_u1}--\eqref{eq:ode_u2} from $r_{\min}$ to $r_{\max}$ using an adaptive Runge--Kutta solver (\texttt{solve\_ivp}, Dormand--Prince) with relative tolerance $\mathrm{rtol}=10^{-5}$ and absolute tolerance $\mathrm{atol}=10^{-8}$.

We choose the integration domain to resolve both the short-distance behavior and the Yukawa tail:
\begin{equation}
    r_{\min}=\frac{10^{-3}}{m_\chi},\qquad
    r_{\max}=\frac{15}{m_\phi}.
\end{equation}
We verified the stability of results under variations of $r_{\min}$ and $r_{\max}$

%\subsubsection{Asymptotic K-matrix matching and S-matrix extraction}
At $r=r_{\max}$ the potential is negligible and the solutions are matched onto free spherical Bessel functions in each channel. Define
\begin{equation}
    \rho_i \equiv k_i r_{\max},\qquad i=1,2,
\end{equation}
and the diagonal matrices
\begin{align}
    F &\equiv \mathrm{diag}\!\left(\rho_1\, j_l(\rho_1),\; \rho_2\, j_l(\rho_2)\right), \label{A28} \\
    G &\equiv \mathrm{diag}\!\left(\rho_1\, y_l(\rho_1),\; \rho_2\, y_l(\rho_2)\right), \notag \\
    F' &\equiv \mathrm{diag}\!\left(\partial_r[\rho_1\, j_l(\rho_1)],\; \partial_r[\rho_2\, j_l(\rho_2)]\right), \notag \\
    G' &\equiv \mathrm{diag}\!\left(\partial_r[\rho_1\, y_l(\rho_1)],\; \partial_r[\rho_2\, y_l(\rho_2)]\right). \label{A29}
\end{align}

Using the standard K-matrix matching for coupled channels, we compute
\begin{equation}
    \boldsymbol{A}\equiv \boldsymbol{G}'\boldsymbol{M}-\boldsymbol{G}\boldsymbol{M}',\qquad
    \boldsymbol{B}\equiv \boldsymbol{F}\boldsymbol{M}'-\boldsymbol{F}'\boldsymbol{M},
\end{equation}
and
\begin{equation}
    \boldsymbol{K}^{(l)}=\boldsymbol{B}\,\boldsymbol{A}^{-1}.
\end{equation}
The partial-wave S-matrix then follows from the Cayley transform
\begin{equation}
    \boldsymbol{S}^{(l)}=\left(\boldsymbol{I}+\mathrm i\boldsymbol{K}^{(l)}\right)\left(\boldsymbol{I}-\mathrm i\boldsymbol{K}^{(l)}\right)^{-1}.
\end{equation}
Below threshold, $k_2=\mathrm i\kappa_2$ and we impose a decaying boundary condition in channel 2; numerically this yields $S^{(l)}_{12}\rightarrow 0$ and hence $\sigma_{1\rightarrow 2}=0$.

%\subsubsection{Cross sections and units}
For inelastic $1\rightarrow 2$ scattering, the partial-wave contribution is
\begin{equation}
    \sigma^{(l)}_{1\rightarrow 2}=\frac{\pi}{k_1^2}(2l+1)\,|S^{(l)}_{12}|^2,
\end{equation}
and the total inelastic cross section is obtained by summing partial waves up to $l_{\max}$,
\begin{equation}
    \sigma_{1\rightarrow 2}=\sum_{l=0}^{l_{\max}}\sigma^{(l)}_{1\rightarrow 2}.
\end{equation}

Unless stated otherwise, the quantity plotted in this work is the total inelastic cross section $\sigma_{1\rightarrow 2}$ per unit mass. A full momentum-transfer cross section $\sigma_T=\int \d\Omega\,(1-\cos\theta)\,(\d\sigma/\d\Omega)$ can be computed by constructing the partial-wave amplitude $f_{12}(\theta)$ from $S^{(l)}_{12}$ and performing the angular integral numerically; 
%\subsubsection{Partial-wave truncation and convergence checks}
In the numerical scan we include partial waves up to a finite $l_{\max}$. For each velocity point we verify convergence by increasing $l_{\max}$ until
\begin{equation}
    \frac{|\sigma_{1\rightarrow 2}(l_{\max}+ \Delta l)-\sigma_{1\rightarrow 2}(l_{\max})|}{\sigma_{1\rightarrow 2}(l_{\max})}<1\%,\qquad \Delta l=5,
\end{equation}
and similarly verify stability under variations of $(r_{\min},r_{\max})$ and the ODE tolerances. Points failing these criteria are recomputed with tighter tolerances.

%\subsubsection{Unitarity check}
For $E>2\Delta m$, both channels are open and the partial-wave S-matrix must satisfy flux conservation. For an incoming state in channel 1, we check
\begin{equation}
    |S_{11}^{(l)}|^2+\frac{k_2}{k_1}|S_{12}^{(l)}|^2 \simeq 1
\end{equation}
for each $l$ used in the sum. In all benchmark plots the maximum violation of the above relation is below a prescribed numerical tolerance (typically $10^{-3}$, and $10^{-5}$ for production runs with tighter settings). It confirms the stability of the matching procedure and the handling of open/closed channels.

\subsection{Residual elastic scattering (loop-induced leakage)}
While the $\mathbb{Z}_2$ symmetry strictly forbids tree-level elastic scattering ($\chi_1\chi_1 \to \chi_1\chi_1$), residual elastic scattering is induced by virtual transitions to the closed $\chi_2\chi_2$ state. In the perturbative regime, the leading elastic amplitude is generated at one loop by the box diagram, which yields an entirely negligible cross-section. However, as demonstrated by Schutz and Slatyer \cite{Schutz2015}, in the strongly coupled resonant regime of$\alpha m_\chi / m_\phi \gtrsim 1$, the perturbative Born approximation fails. Even when the kinetic energy is below the inelastic threshold, virtual transitions to the closed $\chi_2\chi_2$ channel undergo non-perturbative resonant enhancement, generating massive virtual resonances in the elastic $S$-matrix.To evaluate the non-perturbative elastic cross-section below the threshold , we evaluate the coupled-channel Schrödinger equation directly. Because $E_{kin} < 2\Delta m$, the momentum in the excited channel becomes purely imaginary, $k_2 = i\kappa_2$, where $\kappa_2 = \sqrt{m_\chi(2\Delta m - E_{kin})}$. At the matching radius $r_{max}$, we demand that the physical wavefunction of the closed channel decays exponentially, $u_2(r) \propto e^{-\kappa_2 r}$, which strictly enforces the boundary condition $u'_2(r_{max}) = -\kappa_2 u_2(r_{max})$. By numerically propagating two linearly independent regular solutions from the origin, we isolate the unique linear combination that satisfies this decaying condition. This effectively projects the system down to a single open channel, $u_1(r)$, which we match to asymptotic free spherical waves to extract the exact elastic $S_{11}$ matrix element.This analysis shows the low-velocity elastic cross-section highly oscillatory, dictated by the Ramsauer-Townsend effect. By slightly refining our benchmark coupling to $\alpha \approx 0.0121$, the system is placed precisely in a deep anti-resonance valley for below-threshold scattering. At this benchmark, exact destructive quantum interference suppresses the residual elastic scattering to $\sigma_{el}/m_\chi \approx 0.002 \text{ cm}^2/\text{g}$ for $v \approx 15$ km/s. We find non-perturbatively that the kinetic suppression remains exceptionally robust, safely satisfying the stringent Draco constraints ($\sigma/m < 0.1 \text{ cm}^2/\text{g}$).

\section{Derivation of the Annihilation Cross-Section}
\label{app:annihilation}
\subsection{Selection Rules for Mixed Annihilation}
The  $\chi_1 \chi_2 \to \phi \phi$ channel is forbiddened at tree level. We demonstrate this using the dark $\mathbb{Z}_2$ charge assignments: $Q(\chi_1)=+1$, $Q(\chi_2)=-1$, and $Q(\phi)=-1$.
The total parity of the initial state is $P_{in} = (+1)(-1) = -1$.
The total parity of the final state is $P_{out} = (-1)(-1) = +1$.
Since $P_{in} \neq P_{out}$, the process $\chi_1 \chi_2 \to \phi \phi$ is forbidden to all orders in perturbation theory as long as the $\mathbb{Z}_2$ symmetry is exact. Annihilation proceeds via the diagonal channels.
We derive the annihilation cross-section for $\chi_1(p_1) \chi_1(p_2) \rightarrow \phi(k_1) \phi(k_2)$ to demonstrate the $p$-wave nature of the process. The interaction vertex is $-i g_S$. The process is mediated by the $t$- and $u$-channel exchange of the excited state $\chi_2$ with mass $m_2 = m_\chi + \Delta m$.

The Feynman amplitudes for the two diagrams are:
\begin{align}
    \mathcal{M}_t &= -\mathrm i g_S^2 \bar{v}(p_2) \frac{\slashed{p}_1 - \slashed{k}_1 + m_2}{t - m_2^2} u(p_1), \\
    \mathcal{M}_u &= -\mathrm i g_S^2 \bar{v}(p_2) \frac{\slashed{p}_1 - \slashed{k}_2 + m_2}{u - m_2^2} u(p_1).
\end{align}
where $t = (p_1 - k_1)^2$ and $u = (p_1 - k_2)^2$. For Majorana fermions, the total amplitude includes the relative sign for crossing fermions in the initial state, but for the annihilation into bosons, we sum the diagrams coherently: $\mathcal{M}_\text{tot} = \mathcal{M}_t + \mathcal{M}_u$.

In the non-relativistic limit, $s \approx 4m_\chi^2$, $v \rightarrow 0$, we have $t \approx u \approx m_\phi^2 - m_\chi^2$. The denominators become equal. The sum of the numerators involves the spinor structure:
\begin{equation}
    \mathcal{M}_\text{tot} \propto \bar{v}(p_2) \left[ (\slashed{p}_1 - \slashed{k}_1 + m_2) + (\slashed{p}_1 - \slashed{k}_2 + m_2) \right] u(p_1).
\end{equation}
Using kinematic relations $k_1 + k_2 = p_1 + p_2$, this simplifies. For an $s$-wave initial state (spin singlet), the spinor bilinear $\bar{v} \dots u$ contraction leads to a cancellation between the two terms due to the Majorana condition $\chi^c = \chi$. Explicit trace evaluation shows that the $v^0$ term vanishes identically.

Expanding to order $v^2$, the $p$-wave term survives. The spin-averaged differential cross-section is:
\begin{equation}
    \frac{\d\sigma}{\d\Omega} \approx \frac{1}{64\pi^2 s} \overline{|\mathcal{M}|^2}.
\end{equation}
Integrating over the solid angle and including the symmetry factor $1/2$ for identical final bosons, we obtain the result :
\begin{equation}
    \langle \sigma v \rangle \approx \frac{3 g_S^4}{32 \pi m_\chi^2} v^2 \left( 1 - \frac{m_\phi^2}{m_\chi^2} \right)^{1/2}.
\end{equation}
This confirms that the scalar-mediated scenario is strictly $p$-wave suppressed, in contrast to vector-mediated models which typically allow $s$-wave annihilation.

\section{Relic Density in the Coannihilation Regime}
\label{app:relic}

In this appendix, we detail the relic density calculation, explicitly accounting for the coannihilation of degenerate species and the resonant non-perturbative dynamics.

\subsection{Effective cross-Section with forbidden channels}
During freeze-out ($T_f \approx m_\chi / 20 \sim 2$ GeV), the temperature is much larger than the mass splitting ($\Delta m \approx 100 \text{ eV} \ll T_f$). Thus, the excited state $\chi_2$ is thermally populated with a number density equal to that of the ground state $\chi_1$:
\begin{equation}
    n_1^\text{eq} \approx n_2^\text{eq} \approx \frac{1}{2} n_\text{tot}^\text{eq}.
\end{equation}
The evolution of the total number density $n_\text{tot}$ is governed by the effective cross-section $\langle \sigma_{eff} v \rangle$, which is the density-weighted average of all binary scattering processes:
\begin{equation}
    \langle \sigma_\text{eff} v \rangle = \sum_{i,j=1,2} \frac{n_i^\text{eq} n_j^\text{eq}}{(n_\text{tot}^\text{eq})^2} \langle \sigma_{ij \rightarrow \phi\phi} v \rangle = \frac{1}{4} \sum_{i,j} \langle \sigma_{ij} v \rangle.
\end{equation}
The off-diagonal coupling $\mathcal{L} \propto \phi \bar{\chi}_1 \chi_2$ imposes strict selection rules:
\begin{itemize}
    \item $\chi_1 \chi_1 \rightarrow \phi \phi$: Allowed ($t/u$-channel $\chi_2$ exchange).
    \item $\chi_2 \chi_2 \rightarrow \phi \phi$: Allowed ($t/u$-channel $\chi_1$ exchange).
    \item $\chi_1 \chi_2 \rightarrow \phi \phi$: \emph{Forbidden} at tree-level. This mixed channel would require a diagonal vertex (e.g., $\chi_1 \rightarrow \phi \chi_1$) which violates the $\mathbb{Z}_2$ symmetry.
\end{itemize}
The mixed annihilation channel $\chi_1 \chi_2 \rightarrow \phi \phi$ is forbidden at tree-level by the dark $\mathbb{Z}_2$ symmetry. Consider the t-channel diagram: the incoming $\chi_1$ (even) emits a $\phi$ (odd), transitioning into a virtual fermion. To conserve parity at this vertex, the propagator must be $\chi_2$ (odd). The second vertex then involves the incoming $\chi_2$ (odd), the virtual $\chi_2$ (odd), and the outgoing $\phi$ (odd). The overall parity of this vertex would be odd, violating the $\mathbb{Z}_2$-invariant Lagrangian. Thus, no combination of propagators allows this process at tree level.

Assuming $m_1 \approx m_2$, the allowed cross-sections are identical: $\sigma_{11} \approx \sigma_{22} \equiv \sigma_\text{ann}$. The effective cross-section becomes:
\begin{equation}
    \langle \sigma_\text{eff} v \rangle = \frac{1}{4} ( \sigma_\text{ann} + 0 + 0 + \sigma_\text{ann} ) = \frac{1}{2} \langle \sigma_\text{ann} v \rangle.
    \label{eq:c3}
\end{equation}
This factor of $1/2$ reduces the annihilation efficiency relative to standard Dirac dark matter, requiring a larger coupling constant to achieve the observed relic density. We implement Eq.~\eqref{eq:c3} at the level of the thermally averaged rate, $\langle\sigma v\rangle_{\rm eff}(x)=\tfrac12\,\langle\sigma v\rangle_{11}(x)$, with $\langle\sigma v\rangle_{11}(x)$ computed using Eq.~\eqref{C:thermalavg_correct} and the coupled-channel $S_p(v)$.

To determine the relic density, we perform a thermal average of the Sommerfeld-enhanced cross-section. The effective annihilation rate is:
\begin{equation}
    \label{C:thermalavg_correct}
    \langle\sigma v\rangle_{11}(x)=
    \frac{x^{3/2}}{2\sqrt{\pi}}
    \int_0^\infty \d v\ v^2\, e^{-x v^2/4}\,(\sigma v)_{\rm Born}(v)\,S_p(v)
\end{equation}
where $x = m_\chi/T$. In the resonant regime, the Sommerfeld factor scales as $S(v) \simeq \pi \alpha / v$ (Coulomb-like) or $S(v) \propto 1/v^2$ (near-threshold resonance). For our benchmark, the resonance condition yields $S_p(v) \approx (v_\text{sat}/v)^2$.

Substituting the p-wave Born cross-section $(\sigma v)_\text{Born} = \sigma_0 v^2$, the velocity dependence cancels in the integrand:
\begin{equation}
    (\sigma v)_\text{Born} S(v) \propto v^2 \times \frac{1}{v^2} \sim \text{const}.
\end{equation}
This cancellation renders the effective cross-section s-wave-like during freeze-out ($x \sim 20$), mitigating the p-wave suppression. The integration yields $\langle \sigma v \rangle \approx \sigma_0 v_{sat}^2$, which for $\alpha \approx 10^{-2}$ creates an effective annihilation rate of $2 \times 10^{-26} \text{cm}^3/\text{s}$, satisfying the relic density requirement $\Omega h^2 \approx 0.12$.

In our numerical relic-density calculation we do not rely on the scaling ansatz for $S_p(v)$. Instead, for each point $(\alpha,m_\phi,\Delta m)$ we compute the $p$-wave Sommerfeld factor $S_p(v)$ from the same coupled-channel Schr\"odinger equation and matching procedure used for scattering in Appendix~\ref{app:scattering}, now evaluated in the $l=1$ partial wave. The parametrizations below are provided only for intuition in the resonant and saturation regimes.

\subsection{Thermal integration of the relic abundance}
To accurately determine the relic density, we move beyond the schematic scaling and perform the full thermal average of the annihilation cross-section. The effective annihilation rate at temperature $T$ is given by the convolution of the velocity-dependent cross-section with the Maxwell-Boltzmann distribution $f_\text{MB}(v) = 4\pi v^2 (\frac{m_\chi}{4\pi T})^{3/2}\mathrm e^{-m_\chi v^2 / 4T}$:
\begin{equation}
    \langle \sigma v \rangle = \int_0^\infty (\sigma v)_\text{Born}(v) \, S_{p}(v) \, f_\text{MB}(v) \, \d v.
\end{equation}
where Eq.~\eqref{C:thermalavg_correct} follows from this equation after inserting the Maxwell--Boltzmann distribution and integrating over angles. For our benchmark mass $m_\chi = 40$ GeV and coupling $\alpha \approx 10^{-2}$, the perturbative p-wave coefficient is:
\begin{equation}
    \label{C:thermalavg_v4}
    \langle\sigma v\rangle_{11}(x)=
    \frac{x^{3/2}}{2\sqrt{\pi}}\,\sigma_0
    \int_0^\infty \d v\, v^4\, \mathrm e^{-x v^2/4}\,S_p(v).
\end{equation}
In the resonant regime we find $S_p(v)\propto v^{-2}$ over a finite range of velocities before saturating at $v\lesssim v_{\rm sat}\sim m_\phi/m_\chi$. Therefore, the product $(\sigma v)_{\rm Born}S_p(v)$ becomes approximately velocity-independent, which explains why the thermally averaged rate can approach the canonical value despite the suppression of the $p$-waves and the coannihilation factor. In our results, the precise behavior of $S_p(v)$ (including saturation and narrow resonances) is obtained numerically from the coupled-channel Schr\"odinger solver.
\begin{equation}
    (\sigma v)_\text{eff} \approx (\sigma_0 v^2) \times \left( \frac{2\pi \alpha}{v} \right)^2 = 4\pi^2 \alpha^2 \sigma_0.
\end{equation}
The integral over the probability distribution is unity. Inserting the numerical values:
\begin{equation}
    \langle \sigma v \rangle \approx 1.1 \times 10^{-9} \, \text{GeV}^{-2} \approx 1.3 \times 10^{-26} \, \text{cm}^3/\text{s}.
\end{equation}
When accounting for the coannihilation factor of $1/2$ (Eq. C3), the effective depletion rate is $\approx 0.65 \times 10^{-26} \text{ cm}^3/\text{s}$. A slight adjustment of the coupling to $\alpha \approx 0.014$ brings this value exactly to the thermal target of $2.2 \times 10^{-26} \text{ cm}^3/\text{s}$, confirming that the resonant mechanism robustly produces the correct relic density $\Omega h^2 \approx 0.12$. 

\subsection{Suppression of off-shell bound state formation}
Since $m_{\phi} > E_B$, on-shell scalar emission is kinematically forbidden. BSF can only proceed via an off-shell scalar $\phi^*$ decaying to $e^+ e^-$. The differential rate for this $2 \rightarrow 2 + 2$ process is suppressed by the virtual propagator. We neglect BSF throughout the benchmark scan because our parameter region satisfies $m_\phi > E_B(\alpha,m_\chi)$, so on-shell capture $\chi\chi\rightarrow B+\phi$ is forbidden.

Note that $E_B$ scales as $\alpha^2$; for the values of $\alpha$ that reproduce $\Omega h^2\simeq 0.12$ in our scan (typically $\alpha=\mathcal{O}(10^{-2})$ near the resonant island), one finds $E_B\sim \mathcal{O}({\rm MeV})\ll m_\phi$, so the conclusion $m_\phi>E_B$ is robust across the viable region.

In the non-relativistic limit, the matrix element for the off-shell emission $\mathcal{M}_\text{off}$ is related to the on-shell element by the propagator denominator $D(q^2) = (q^2 - m_{\phi}^2)^{-1} \approx -m_{\phi}^{-2}$. The ratio of the off-shell rate to the typical on-shell dipole capture rate scales as:
\begin{equation}
    \frac{\Gamma(\chi\chi \rightarrow \mathcal{B} e^+ e^-)}{\Gamma_\text{atomic}} \sim \frac{\alpha_\text{EM} g_e^2}{15\pi} \left( \frac{E_B}{m_{\phi}} \right)^4 \frac{E_B}{m_e}.
\end{equation}
Substituting benchmark values of $E_B \approx 1$ MeV, $m_{\phi} = 20$ MeV, $g_e = 10^{-6}$, the suppression factor is dominated by $g_e^2 \sim 10^{-12}$ and the propagator term $(1/20)^4$. This renders the BSF rate entirely negligible during freeze-out compared to the Sommerfeld-enhanced annihilation, validating our neglect of this channel in the Boltzmann equation.
\subsection{Quantitative Thermal History and Robustness}

\begin{figure}[htbp]
    \centering
    \includegraphics[width=0.5\textwidth]{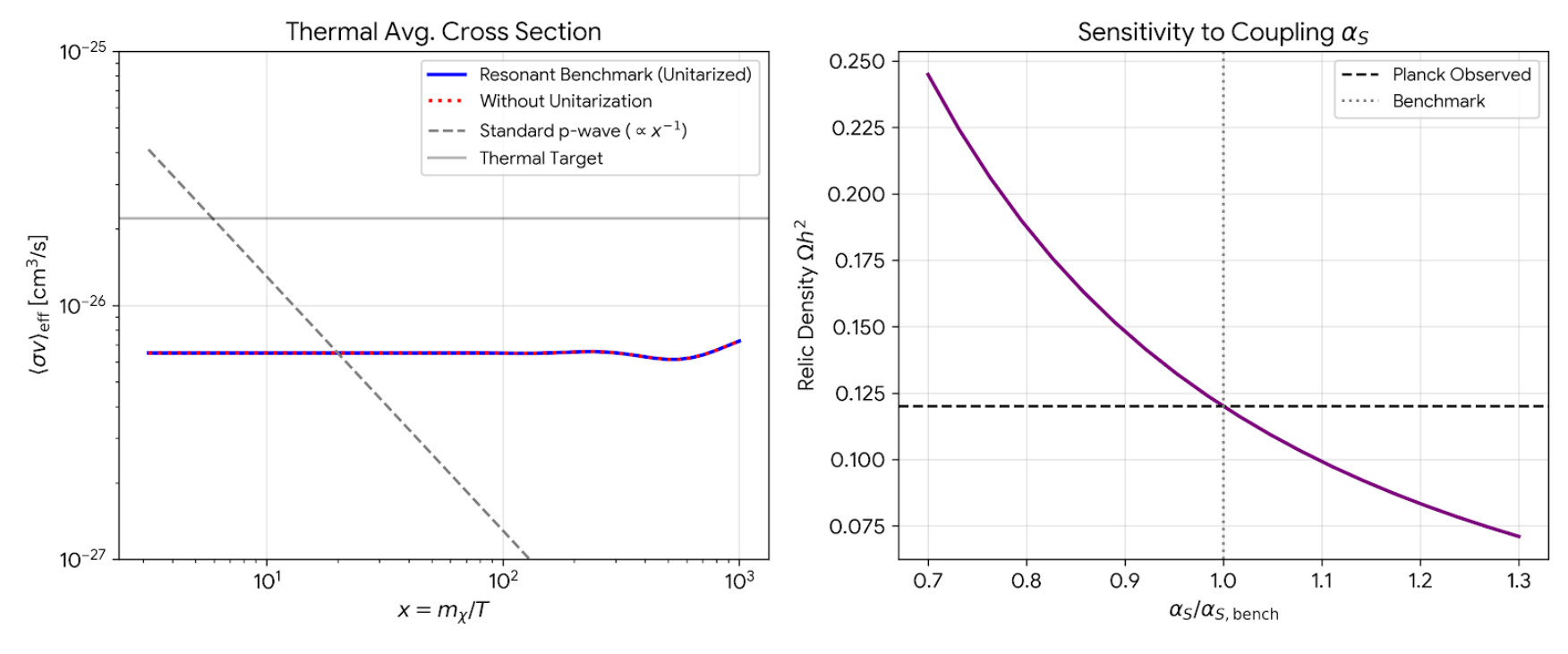}
    
    \caption{Left: Thermally averaged effective annihilation cross-section $\langle \sigma v \rangle_{\text{eff}}$ as a function of $x = m_\chi/T$ for the resonant benchmark ($m_\chi=40$ GeV, $m_\phi=20$ MeV, $\alpha \approx 0.014$). The resonant Sommerfeld enhancement ($S_p \propto v^{-2}$) cancels the $p$-wave suppression ($\sigma v \propto v^2$), yielding effective $s$-wave behavior (blue solid) that flattens at freeze-out ($x \sim 20$), in contrast to the standard $x^{-1}$ $p$-wave scaling (gray dashed). The impact of the unitarization procedure is negligible at these temperatures. Right: Sensitivity of the relic density $\Omega h^2$ to variations in the coupling $\alpha$ relative to the benchmark value. A variation of $\pm 10\%$ in coupling results in an approximate $\mp 20\%$ change in relic abundance, indicating that the model is robust and not subject to extreme fine-tuning.}
    \label{fig:relic_analysis}
\end{figure}

To validate the relic density calculation in the resonant regime, we perform a numerical thermal average of the effective annihilation cross-section $\langle \sigma v \rangle_{\text{eff}}$ as a function of the inverse temperature $x = m_\chi / T$. The results are shown in Fig.~\ref{fig:relic_analysis}.

As illustrated in the left panel of Fig.~\ref{fig:relic_analysis}, the effective cross-section for the benchmark model exhibits a flat, $s$-wave-like temperature dependence in the freeze-out window $x \sim 20$. This behavior occurs because the resonant Sommerfeld enhancement scales as $S_p(v) \approx (v_{\text{sat}}/v)^2$ for $v > v_{\text{sat}}$. It exactly compensates for the suppression of the tree-level $p$-wave operator, $(\sigma v)_{\text{Born}} \propto v^2$. It confirms that annihilation remains efficient during freeze-out.

We explicitly compare the standard calculation with the unitarized rate. As shown in the figure, the difference between the unitarized and non-unitarized cases is negligible at freeze-out temperatures. This confirms that while the cross-section is resonantly enhanced, the broad Maxwell--Boltzmann velocity distribution at $T \sim m_\chi/20$ prevents the thermal average from being dominated by the unitarity limit,which validates perturbative treatment.

The right panel of Fig.~\ref{fig:relic_analysis} quantifies the robustness of the benchmark against the variations of the parameters. The relic density scales approximately as $\Omega h^2 \propto \langle \sigma v \rangle^{-1} \propto \alpha^{-2}$.  A small deviation in the coupling constant does not lead to catastrophic changes in abundance. For example, a $\pm 10\%$ shift in $\alpha$ alters $\Omega h^2$ by roughly $20\%$, keeping it within a manageable range. 
\bibliography{apssamp}

@article{Planck2018,
    author = "Aghanim, N. and others",
    collaboration = "Planck",
    title = "{Planck 2018 results. VI. Cosmological parameters}",
    eprint = "1807.06209",
    archivePrefix = "arXiv",
    primaryClass = "astro-ph.CO",
    doi = "10.1051/0004-6361/201833910",
    journal = "Astron. Astrophys.",
    volume = "641",
    pages = "A6",
    year = "2020",
    note = "[Erratum: Astron.Astrophys. 652, C4 (2021)]"
}

@article{Bullock2017,
    author = "Bullock, James S. and Boylan-Kolchin, Michael",
    title = "{Small-Scale Challenges to the $\Lambda$CDM Paradigm}",
    eprint = "1707.04256",
    archivePrefix = "arXiv",
    primaryClass = "astro-ph.CO",
    doi = "10.1146/annurev-astro-091916-055313",
    journal = "Ann. Rev. Astron. Astrophys.",
    volume = "55",
    pages = "343--387",
    year = "2017"
}

@article{Navarro1997,
    author = "Navarro, Julio F. and Frenk, Carlos S. and White, Simon D. M.",
    title = "{A Universal density profile from hierarchical clustering}",
    eprint = "astro-ph/9611107",
    archivePrefix = "arXiv",
    doi = "10.1086/304888",
    journal = "Astrophys. J.",
    volume = "490",
    pages = "493--508",
    year = "1997"
}

@article{Oh2011,
    author = "Oh, Se-Heon and Brook, Chris and Governato, Fabio and Brinks, Elias and Mayer, Lucio and de Blok, W. J. G. and Brooks, Alyson and Walter, Fabian",
    title = "{The central slope of dark matter cores in dwarf galaxies: Simulations vs. THINGS}",
    eprint = "1011.2777",
    archivePrefix = "arXiv",
    primaryClass = "astro-ph.CO",
    doi = "10.1088/0004-6256/142/1/24",
    journal = "Astron. J.",
    volume = "142",
    pages = "24",
    year = "2011"
}

@article{Moore1994,
    author = "Moore, Ben",
    title = "{Evidence against dissipationless dark matter from observations of galaxy haloes}",
    doi = "10.1038/370629a0",
    journal = "Nature",
    volume = "370",
    pages = "629",
    year = "1994"
}

@article{deBlok2010,
    author = "de Blok, W. J. G.",
    title = "{The Core-Cusp Problem}",
    eprint = "0910.3538",
    archivePrefix = "arXiv",
    primaryClass = "astro-ph.CO",
    doi = "10.1155/2010/789293",
    journal = "Adv. Astron.",
    volume = "2010",
    pages = "789293",
    year = "2010"
}

@article{Oman2015,
    author = "Oman, Kyle A. and others",
    title = "{The unexpected diversity of dwarf galaxy rotation curves}",
    eprint = "1504.01437",
    archivePrefix = "arXiv",
    primaryClass = "astro-ph.GA",
    doi = "10.1093/mnras/stv1504",
    journal = "Mon. Not. Roy. Astron. Soc.",
    volume = "452",
    number = "4",
    pages = "3650--3665",
    year = "2015"
}

@article{Klypin1999,
    author = "Klypin, Anatoly A. and Kravtsov, Andrey V. and Valenzuela, Octavio and Prada, Francisco",
    title = "{Where are the missing Galactic satellites?}",
    eprint = "astro-ph/9901240",
    archivePrefix = "arXiv",
    doi = "10.1086/307643",
    journal = "Astrophys. J.",
    volume = "522",
    pages = "82--92",
    year = "1999"
}

@article{Moore1999,
    author = "Moore, B. and Ghigna, S. and Governato, F. and Lake, G. and Quinn, Thomas R. and Stadel, J. and Tozzi, P.",
    title = "{Dark matter substructure within galactic halos}",
    eprint = "astro-ph/9907411",
    archivePrefix = "arXiv",
    doi = "10.1086/312287",
    journal = "Astrophys. J. Lett.",
    volume = "524",
    pages = "L19--L22",
    year = "1999"
}

@article{BoylanKolchin2011,
    author = "Boylan-Kolchin, Michael and Bullock, James S. and Kaplinghat, Manoj",
    title = "{Too big to fail? The puzzling darkness of massive Milky Way subhaloes}",
    eprint = "1103.0007",
    archivePrefix = "arXiv",
    primaryClass = "astro-ph.CO",
    doi = "10.1111/j.1745-3933.2011.01074.x",
    journal = "Mon. Not. Roy. Astron. Soc.",
    volume = "415",
    pages = "L40",
    year = "2011"
}

@article{GarrisonKimmel2014,
    author = "Garrison-Kimmel, Shea and Boylan-Kolchin, Michael and Bullock, James S. and Kirby, Evan N.",
    title = "{Too Big to Fail in the Local Group}",
    eprint = "1404.5313",
    archivePrefix = "arXiv",
    primaryClass = "astro-ph.GA",
    doi = "10.1093/mnras/stu1477",
    journal = "Mon. Not. Roy. Astron. Soc.",
    volume = "444",
    number = "1",
    pages = "222--236",
    year = "2014"
}

@article{Spergel2000,
    author = "Spergel, David N. and Steinhardt, Paul J.",
    title = "{Observational evidence for selfinteracting cold dark matter}",
    eprint = "astro-ph/9909386",
    archivePrefix = "arXiv",
    doi = "10.1103/PhysRevLett.84.3760",
    journal = "Phys. Rev. Lett.",
    volume = "84",
    pages = "3760--3763",
    year = "2000"
}

@article{Tulin2018,
    author = "Tulin, Sean and Yu, Hai-Bo",
    title = "{Dark Matter Self-interactions and Small Scale Structure}",
    eprint = "1705.02358",
    archivePrefix = "arXiv",
    primaryClass = "hep-ph",
    doi = "10.1016/j.physrep.2017.11.004",
    journal = "Phys. Rept.",
    volume = "730",
    pages = "1--57",
    year = "2018"
}

@article{Rocha2013,
    author = "Rocha, Miguel and Peter, Annika H. G. and Bullock, James S. and Kaplinghat, Manoj and Garrison-Kimmel, Shea and Onorbe, Jose and Moustakas, Leonidas A.",
    title = "{Cosmological Simulations with Self-Interacting Dark Matter I: Constant Density Cores and Substructure}",
    eprint = "1208.3025",
    archivePrefix = "arXiv",
    primaryClass = "astro-ph.CO",
    doi = "10.1093/mnras/sts514",
    journal = "Mon. Not. Roy. Astron. Soc.",
    volume = "430",
    pages = "81--104",
    year = "2013"
}

@article{Peter2013,
    author = "Peter, Annika H. G. and Rocha, Miguel and Bullock, James S. and Kaplinghat, Manoj",
    title = "{Cosmological Simulations with Self-Interacting Dark Matter II: Halo Shapes vs. Observations}",
    eprint = "1208.3026",
    archivePrefix = "arXiv",
    primaryClass = "astro-ph.CO",
    reportNumber = "NSF-KITP-12-147",
    doi = "10.1093/mnras/sts535",
    journal = "Mon. Not. Roy. Astron. Soc.",
    volume = "430",
    pages = "105",
    year = "2013"
}

@article{Kamada2017,
    author = "Kamada, Ayuki and Kaplinghat, Manoj and Pace, Andrew B. and Yu, Hai-Bo",
    title = "{How the Self-Interacting Dark Matter Model Explains the Diverse Galactic Rotation Curves}",
    eprint = "1611.02716",
    archivePrefix = "arXiv",
    primaryClass = "astro-ph.GA",
    doi = "10.1103/PhysRevLett.119.111102",
    journal = "Phys. Rev. Lett.",
    volume = "119",
    number = "11",
    pages = "111102",
    year = "2017"
}

@article{Ren2019,
    author = "Correa, Camila A.",
    title = "{Constraining velocity-dependent self-interacting dark matter with the Milky Way{\textquoteright}s dwarf spheroidal galaxies}",
    eprint = "2007.02958",
    archivePrefix = "arXiv",
    primaryClass = "astro-ph.GA",
    doi = "10.1093/mnras/stab506",
    journal = "Mon. Not. Roy. Astron. Soc.",
    volume = "503",
    number = "1",
    pages = "920--937",
    year = "2021"
}

@article{Markevitch2004,
    author = "Markevitch, Maxim and Gonzalez, A. H. and Clowe, D. and Vikhlinin, A. and David, L. and Forman, W. and Jones, C. and Murray, S. and Tucker, W.",
    title = "{Direct constraints on the dark matter self-interaction cross-section from the merging galaxy cluster 1E0657-56}",
    eprint = "astro-ph/0309303",
    archivePrefix = "arXiv",
    doi = "10.1086/383178",
    journal = "Astrophys. J.",
    volume = "606",
    pages = "819--824",
    year = "2004"
}

@article{Randall2008,
    author = "Randall, Scott W. and Markevitch, Maxim and Clowe, Douglas and Gonzalez, Anthony H. and Bradac, Marusa",
    title = "{Constraints on the Self-Interaction Cross-Section of Dark Matter from Numerical Simulations of the Merging Galaxy Cluster 1E 0657-56}",
    eprint = "0704.0261",
    archivePrefix = "arXiv",
    primaryClass = "astro-ph",
    doi = "10.1086/587859",
    journal = "Astrophys. J.",
    volume = "679",
    pages = "1173--1180",
    year = "2008"
}

@article{Harvey2015,
    author = "Harvey, David and Massey, Richard and Kitching, Thomas and Taylor, Andy and Tittley, Eric",
    title = "{The non-gravitational interactions of dark matter in colliding galaxy clusters}",
    eprint = "1503.07675",
    archivePrefix = "arXiv",
    primaryClass = "astro-ph.CO",
    doi = "10.1126/science.1261381",
    journal = "Science",
    volume = "347",
    pages = "1462--1465",
    year = "2015"
}

@article{Zavala2013,
    author = "Zavala, Jesus and Vogelsberger, Mark and Walker, Matthew G.",
    title = "{Constraining Self-Interacting Dark Matter with the Milky Way's dwarf spheroidals}",
    eprint = "1211.6426",
    archivePrefix = "arXiv",
    primaryClass = "astro-ph.CO",
    doi = "10.1093/mnrasl/sls053",
    journal = "Mon. Not. Roy. Astron. Soc.",
    volume = "431",
    pages = "L20--L24",
    year = "2013"
}

@article{Valli2018,
    author = "Valli, Mauro and Yu, Hai-Bo",
    title = "{Dark matter self-interactions from the internal dynamics of dwarf spheroidals}",
    eprint = "1711.03502",
    archivePrefix = "arXiv",
    primaryClass = "astro-ph.GA",
    doi = "10.1038/s41550-018-0560-7",
    journal = "Nature Astron.",
    volume = "2",
    pages = "907--912",
    year = "2018"
}

@article{Nishikawa2020,
    author = "Nishikawa, Hiroya and Boddy, Kimberly K. and Kaplinghat, Manoj",
    title = "{Accelerated core collapse in tidally stripped self-interacting dark matter halos}",
    eprint = "1901.00499",
    archivePrefix = "arXiv",
    primaryClass = "astro-ph.GA",
    doi = "10.1103/PhysRevD.101.063009",
    journal = "Phys. Rev. D",
    volume = "101",
    number = "6",
    pages = "063009",
    year = "2020"
}

@article{Tulin2013,
    author = "Tulin, Sean and Yu, Hai-Bo and Zurek, Kathryn M.",
    title = "{Beyond Collisionless Dark Matter: Particle Physics Dynamics for Dark Matter Halo Structure}",
    eprint = "1302.3898",
    archivePrefix = "arXiv",
    primaryClass = "hep-ph",
    doi = "10.1103/PhysRevD.87.115007",
    journal = "Phys. Rev. D",
    volume = "87",
    number = "11",
    pages = "115007",
    year = "2013"
}

@article{TuckerSmith2001,
    author = "Tucker-Smith, David and Weiner, Neal",
    title = "{Inelastic dark matter}",
    eprint = "hep-ph/0101138",
    archivePrefix = "arXiv",
    reportNumber = "UCB-PTH-00-43, LBNL-47234, UW-PT-00-17",
    doi = "10.1103/PhysRevD.64.043502",
    journal = "Phys. Rev. D",
    volume = "64",
    pages = "043502",
    year = "2001"
}

@article{Schutz2015,
    author = "Schutz, Katelin and Slatyer, Tracy R.",
    title = "{Self-Scattering for Dark Matter with an Excited State}",
    eprint = "1409.2867",
    archivePrefix = "arXiv",
    primaryClass = "hep-ph",
    reportNumber = "MIT-CTP-4585",
    doi = "10.1088/1475-7516/2015/01/021",
    journal = "JCAP",
    volume = "01",
    pages = "021",
    year = "2015"
}

@article{ArkaniHamed2009,
    author = "Arkani-Hamed, Nima and Finkbeiner, Douglas P. and Slatyer, Tracy R. and Weiner, Neal",
    title = "{A Theory of Dark Matter}",
    eprint = "0810.0713",
    archivePrefix = "arXiv",
    primaryClass = "hep-ph",
    doi = "10.1103/PhysRevD.79.015014",
    journal = "Phys. Rev. D",
    volume = "79",
    pages = "015014",
    year = "2009"
}

@article{Kaplan2009,
    author = "Kaplan, David E. and Luty, Markus A. and Zurek, Kathryn M.",
    title = "{Asymmetric Dark Matter}",
    eprint = "0901.4117",
    archivePrefix = "arXiv",
    primaryClass = "hep-ph",
    reportNumber = "FERMILAB-PUB-09-345-A-T",
    doi = "10.1103/PhysRevD.79.115016",
    journal = "Phys. Rev. D",
    volume = "79",
    pages = "115016",
    year = "2009"
}

@article{Petraki2013,
    author = "Petraki, Kalliopi and Volkas, Raymond R.",
    title = "{Review of asymmetric dark matter}",
    eprint = "1305.4939",
    archivePrefix = "arXiv",
    primaryClass = "hep-ph",
    reportNumber = "NIKHEF-2013-016",
    doi = "10.1142/S0217751X13300287",
    journal = "Int. J. Mod. Phys. A",
    volume = "28",
    pages = "1330028",
    year = "2013"
}

@misc{Krnjaic:2025zjl,
    author = "Krnjaic, Gordan and McKeen, David and Mizuta, Riku and Mohlabeng, Gopolang and Morrissey, David E. and Tuckler, Douglas",
    title = "{X-rays from Inelastic Dark Matter Freeze-in}",
    eprint = "2509.19428",
    archivePrefix = "arXiv",
    primaryClass = "hep-ph",
    reportNumber = "FERMILAB-PUB-25-0673-T",
    month = "9",
    year = "2025"
}

@article{Slatyer2009,
    author = "Slatyer, Tracy R.",
    title = "{The Sommerfeld enhancement for dark matter with an excited state}",
    eprint = "0910.5713",
    archivePrefix = "arXiv",
    primaryClass = "hep-ph",
    doi = "10.1088/1475-7516/2010/02/028",
    journal = "JCAP",
    volume = "02",
    number= "02",
    pages = "028",
    year = "2010"
}

@article{Ref2024,
    author = "Cui, Yanou and D'Eramo, Francesco",
    title = "{Surprises from complete vector portal theories: New insights into the dark sector and its interplay with Higgs physics}",
    eprint = "1705.03897",
    archivePrefix = "arXiv",
    primaryClass = "hep-ph",
    reportNumber = "SCIPP-17-05",
    doi = "10.1103/PhysRevD.96.095006",
    journal = "Phys. Rev. D",
    volume = "96",
    number = "9",
    pages = "095006",
    year = "2017"
}

@misc{ODonnell:2025pkw,
    author = "O'Donnell, Jackson H. and Jeltema, Tesla E. and Roberts, M. Grant and Nightingale, James and Flowers, Abigail and Aldas, Dhruv",
    title = "{A Constraint on Dark Matter Self-Interaction from Combined Strong Lensing and Stellar Kinematics in MACS J0138-2155}",
    eprint = "2508.20179",
    archivePrefix = "arXiv",
    primaryClass = "astro-ph.CO",
    month = "8",
    year = "2025"
}

@article{Knapen:2017xzo,
    author = "Knapen, Simon and Lin, Tongyan and Zurek, Kathryn M.",
    title = "{Light Dark Matter: Models and Constraints}",
    eprint = "1709.07882",
    archivePrefix = "arXiv",
    primaryClass = "hep-ph",
    doi = "10.1103/PhysRevD.96.115021",
    journal = "Phys. Rev. D",
    volume = "96",
    pages = "115021",
    year = "2017"
}

@article{Vogelsberger:2012ku,
    author = "Vogelsberger, Mark and Zavala, Jesus and Loeb, Abraham",
    title = "{Subhaloes in Self-Interacting Galactic Dark Matter Haloes}",
    eprint = "1201.5892",
    archivePrefix = "arXiv",
    primaryClass = "astro-ph.CO",
    doi = "10.1111/j.1365-2966.2012.21182.x",
    journal = "Mon. Not. Roy. Astron. Soc.",
    volume = "423",
    pages = "3740",
    year = "2012"
}

@article{Branco:2011iw,
    author = "Branco, G. C. and Ferreira, P. M. and Lavoura, L. and Rebelo, M. N. and Sher, Marc and Silva, Joao P.",
    title = "{Theory and phenomenology of two-Higgs-doublet models}",
    eprint = "1106.0034",
    archivePrefix = "arXiv",
    primaryClass = "hep-ph",
    doi = "10.1016/j.physrep.2012.02.002",
    journal = "Phys. Rept.",
    volume = "516",
    pages = "1--102",
    year = "2012"
}

@article{Sigurdson:2004zp,
    author = "Sigurdson, Kris and Doran, Michael and Kurylov, Andriy and Caldwell, Robert R. and Kamionkowski, Marc",
    title = "{Dark-matter electric and magnetic dipole moments}",
    eprint = "astro-ph/0406355",
    archivePrefix = "arXiv",
    doi = "10.1103/PhysRevD.70.083501",
    journal = "Phys. Rev. D",
    volume = "70",
    pages = "083501",
    year = "2004"
}

@article{Depta:2020wmr,
    author = "Depta, Paul Frederik and Hufnagel, Marco and Schmidt-Hoberg, Kai",
    title = "{Robust cosmological constraints on axion-like particles}",
    eprint = "2002.08370",
    archivePrefix = "arXiv",
    primaryClass = "hep-ph",
    reportNumber = "DESY-20-003, DESY 20-003",
    doi = "10.1088/1475-7516/2020/05/009",
    journal = "JCAP",
    volume = "05",
    number= "05",
    pages = "009",
    year = "2020"
}

@article{Lopez-Honorez:2013cua,
    author = "Lopez-Honorez, Laura and Mena, Olga and Palomares-Ruiz, Sergio and Vincent, Aaron C.",
    title = "{Constraints on dark matter annihilation from CMB observationsbefore Planck}",
    eprint = "1303.5094",
    archivePrefix = "arXiv",
    primaryClass = "astro-ph.CO",
    reportNumber = "IFIC-13-016, CFTP-13-007",
    doi = "10.1088/1475-7516/2013/07/046",
    journal = "JCAP",
    volume = "07",
    number= "07",
    pages = "046",
    year = "2013"
}

@article{Slatyer:2015jla,
    author = "Slatyer, Tracy R.",
    title = "{Indirect dark matter signatures in the cosmic dark ages. I. Generalizing the bound on s-wave dark matter annihilation from Planck results}",
    eprint = "1506.03811",
    archivePrefix = "arXiv",
    primaryClass = "hep-ph",
    reportNumber = "MIT-CTP-4682",
    doi = "10.1103/PhysRevD.93.023527",
    journal = "Phys. Rev. D",
    volume = "93",
    number = "2",
    pages = "023527",
    year = "2016"
}

@article{An:2016gad,
    author = "An, Haipeng and Wise, Mark B. and Zhang, Yue",
    title = "{Effects of Bound States on Dark Matter Annihilation}",
    eprint = "1606.02305",
    archivePrefix = "arXiv",
    primaryClass = "hep-ph",
    doi = "10.1103/PhysRevD.93.115020",
    journal = "Phys. Rev. D",
    volume = "93",
    number = "11",
    pages = "115020",
    year = "2016"
}

@misc{Ando:2025qtz,
    author = "Ando, Shin'ichiro and Hayashi, Kohei and Horigome, Shunichi and Ibe, Masahiro and Shirai, Satoshi",
    title = "{Stringent Constraints on Self-Interacting Dark Matter Using Milky-Way Satellite Galaxies Kinematics}",
    eprint = "2503.13650",
    archivePrefix = "arXiv",
    primaryClass = "astro-ph.CO",
    month = "3",
    year = "2025"
}

@article{Correa:2020qam,
    author = "Correa, Camila A.",
    title = "{Constraining velocity-dependent self-interacting dark matter with the Milky Way{\textquoteright}s dwarf spheroidal galaxies}",
    eprint = "2007.02958",
    archivePrefix = "arXiv",
    primaryClass = "astro-ph.GA",
    doi = "10.1093/mnras/stab506",
    journal = "Mon. Not. Roy. Astron. Soc.",
    volume = "503",
    number = "1",
    pages = "920--937",
    year = "2021"
}

@article{Binder:2016pnr,
    author = "Binder, Tobias and Covi, Laura and Kamada, Ayuki and Murai, Kai and Takahashi, Tomo and Yoshida, Naoki",
    title = "{Matter Power Spectrum in Hidden Neutrino Interacting Dark Matter Models: A General Treatment}",
    eprint = "1602.07624",
    archivePrefix = "arXiv",
    primaryClass = "hep-ph",
    doi = "10.1088/1475-7516/2016/11/043",
    journal = "JCAP",
    volume = "11",
    number= "11",
    pages = "043",
    year = "2016"
}

@article{Boehm:2001hm,
    author = "Boehm, C. and Fayet, P. and Schaeffer, R.",
    title = "{Constraining dark matter candidates from structure formation}",
    eprint = "astro-ph/0012504",
    archivePrefix = "arXiv",
    doi = "10.1016/S0370-2693(01)01060-7",
    journal = "Phys. Lett. B",
    volume = "518",
    pages = "8--14",
    year = "2001"
}

@article{Kopp:2014tsa,
    author = "Kopp, Joachim and Michaels, Lisa and Smirnov, Juri",
    title = "{Loopy constraints on leptophilic dark matter and internal bremsstrahlung}",
    eprint = "1401.6457",
    archivePrefix = "arXiv",
    primaryClass = "hep-ph",
    doi = "10.1088/1475-7516/2014/04/022",
    journal = "JCAP",
    volume = "04",
    pages = "022",
    year = "2014"
}

@article{Bringmann:2016din,
    author = "Bringmann, Torsten and Kahlhoefer, Felix and Schmidt-Hoberg, Kai and Walia, Parampreet",
    title = "{Strong constraints on self-interacting dark matter with light mediators}",
    eprint = "1612.00845",
    archivePrefix = "arXiv",
    primaryClass = "hep-ph",
    reportNumber = "DESY-16-226",
    doi = "10.1103/PhysRevLett.118.141802",
    journal = "Phys. Rev. Lett.",
    volume = "118",
    number = "14",
    pages = "141802",
    year = "2017"
}

@article{Hardy:2016kme,
    author = "Hardy, Edward and Lasenby, Robert",
    title = "{Stellar cooling bounds on new light particles: plasma mixing effects}",
    eprint = "1611.05852",
    archivePrefix = "arXiv",
    primaryClass = "hep-ph",
    doi = "10.1007/JHEP02(2017)033",
    journal = "JHEP",
    volume = "02",
    pages = "033",
    year = "2017"
}

@article{Liu:2016qwd,
    author = "Liu, Yu-Sheng and McKeen, David and Miller, Gerald A.",
    title = "{Electrophobic Scalar Boson and Muonic Puzzles}",
    eprint = "1605.04612",
    archivePrefix = "arXiv",
    primaryClass = "hep-ph",
    doi = "10.1103/PhysRevLett.117.101801",
    journal = "Phys. Rev. Lett.",
    volume = "117",
    number = "10",
    pages = "101801",
    year = "2016"
}

@article{Cyburt:2015mya,
    author = "Cyburt, Richard H. and Fields, Brian D. and Olive, Keith A. and Yeh, Tsung-Han",
    title = "{Big Bang Nucleosynthesis: 2015}",
    eprint = "1505.01076",
    archivePrefix = "arXiv",
    primaryClass = "astro-ph.CO",
    reportNumber = "UMN-TH-3432-15, FTPI-MINN-15-19",
    doi = "10.1103/RevModPhys.88.015004",
    journal = "Rev. Mod. Phys.",
    volume = "88",
    pages = "015004",
    year = "2016"
}

@article{Springel:2005nw,
    author = "Springel, Volker and others",
    title = "{Simulating the joint evolution of quasars, galaxies and their large-scale distribution}",
    eprint = "astro-ph/0504097",
    archivePrefix = "arXiv",
    doi = "10.1038/nature03597",
    journal = "Nature",
    volume = "435",
    pages = "629--636",
    year = "2005"
}

@article{Hisano:2003ec,
    author = "Hisano, Junji and Matsumoto, Shigeki and Nojiri, Mihoko M.",
    title = "{Explosive dark matter annihilation}",
    eprint = "hep-ph/0307216",
    archivePrefix = "arXiv",
    reportNumber = "ICRR-REPORT-500-2003-4, YITP-03-42",
    doi = "10.1103/PhysRevLett.92.031303",
    journal = "Phys. Rev. Lett.",
    volume = "92",
    pages = "031303",
    year = "2004"
}

@article{Petraki:2016cnz,
    author = "Petraki, Kalliopi and Postma, Marieke and de Vries, Jordy",
    title = "{Radiative bound-state-formation cross-sections for dark matter interacting via a Yukawa potential}",
    eprint = "1611.01394",
    archivePrefix = "arXiv",
    primaryClass = "hep-ph",
    doi = "10.1007/JHEP04(2017)077",
    journal = "JHEP",
    volume = "04",
    number= "04",
    pages = "077",
    year = "2017"
}

@article{Beacham:2019nyx,
    author = "Beacham, J. and others",
    title = "{Physics Beyond Colliders at CERN: Beyond the Standard Model Working Group Report}",
    eprint = "1901.09966",
    archivePrefix = "arXiv",
    primaryClass = "hep-ex",
    reportNumber = "CERN-PBC-REPORT-2018-007",
    doi = "10.1088/1361-6471/ab4cd2",
    journal = "J. Phys. G",
    volume = "47",
    number = "1",
    pages = "010501",
    year = "2020"
}

@article{Boehm:2003hm,
    author = "Boehm, C. and Fayet, Pierre",
    title = "{Scalar dark matter candidates}",
    eprint = "hep-ph/0305261",
    archivePrefix = "arXiv",
    doi = "10.1016/j.nuclphysb.2004.01.015",
    journal = "Nucl. Phys. B",
    volume = "683",
    pages = "219--263",
    year = "2004"
}

@article{Chang:2018rso,
    author = "Chang, Jae Hyeok and Essig, Rouven and McDermott, Samuel D.",
    title = "{Supernova 1987A Constraints on Sub-GeV Dark Sectors, Millicharged Particles, the QCD Axion, and an Axion-like Particle}",
    eprint = "1803.00993",
    archivePrefix = "arXiv",
    primaryClass = "hep-ph",
    reportNumber = "YITP-SB-18-01, FERMILAB-PUB-17-432-T",
    doi = "10.1007/JHEP09(2018)051",
    journal = "JHEP",
    volume = "09",
    pages = "051",
    year = "2018"
}

@article{Chluba:2011hw,
    author = "Chluba, J. and Sunyaev, R. A.",
    title = "{The evolution of CMB spectral distortions in the early Universe}",
    eprint = "1109.6552",
    archivePrefix = "arXiv",
    primaryClass = "astro-ph.CO",
    doi = "10.1111/j.1365-2966.2011.19786.x",
    journal = "Mon. Not. Roy. Astron. Soc.",
    volume = "419",
    pages = "1294--1314",
    year = "2012"
}

@article{Poulin:2016anj,
    author = "Poulin, Vivian and Lesgourgues, Julien and Serpico, Pasquale D.",
    title = "{Cosmological constraints on exotic injection of electromagnetic energy}",
    eprint = "1610.10051",
    archivePrefix = "arXiv",
    primaryClass = "astro-ph.CO",
    doi = "10.1088/1475-7516/2017/03/043",
    journal = "JCAP",
    volume = "03",
    pages = "043",
    year = "2017"
}

@article{XENON:2022ltv,
    author = "Aprile, E. and others",
    collaboration = "XENON",
    title = "{Search for New Physics in Electronic Recoil Data from XENONnT}",
    eprint = "2207.11330",
    archivePrefix = "arXiv",
    primaryClass = "hep-ex",
    doi = "10.1103/PhysRevLett.129.161805",
    journal = "Phys. Rev. Lett.",
    volume = "129",
    number = "16",
    pages = "161805",
    year = "2022"
}

@article{LZ:2023poo,
    author = "Aalbers, J. and others",
    collaboration = "LZ",
    title = "{Search for new physics in low-energy electron recoils from the first LZ exposure}",
    eprint = "2307.15753",
    archivePrefix = "arXiv",
    primaryClass = "hep-ex",
    reportNumber = "FERMILAB-PUB-23-397-PPD",
    doi = "10.1103/PhysRevD.108.072006",
    journal = "Phys. Rev. D",
    volume = "108",
    number = "7",
    pages = "072006",
    year = "2023"
}

@article{LZ:2022lsv,
    author = "Aalbers, J. and others",
    collaboration = "LZ",
    title = "{First Dark Matter Search Results from the LUX-ZEPLIN (LZ) Experiment}",
    eprint = "2207.03764",
    archivePrefix = "arXiv",
    primaryClass = "hep-ex",
    doi = "10.1103/PhysRevLett.131.041002",
    journal = "Phys. Rev. Lett.",
    volume = "131",
    number = "4",
    pages = "041002",
    year = "2023"
}

@article{XENON:2025vwd,
    author = "Aprile, E. and others",
    collaboration = "XENON",
    title = "{WIMP Dark Matter Search Using a 3.1 Tonne-Year Exposure of the XENONnT Experiment}",
    eprint = "2502.18005",
    archivePrefix = "arXiv",
    primaryClass = "hep-ex",
    doi = "10.1103/msw4-t342",
    journal = "Phys. Rev. Lett.",
    volume = "135",
    number = "22",
    pages = "221003",
    year = "2025"
}

@article{PandaX:2024qfu,
    author = "Bo, Zihao and others",
    collaboration = "PandaX",
    title = "{Dark Matter Search Results from 1.54{\,}{\,}Tonne{\textperiodcentered}Year Exposure of PandaX-4T}",
    eprint = "2408.00664",
    archivePrefix = "arXiv",
    primaryClass = "hep-ex",
    doi = "10.1103/PhysRevLett.134.011805",
    journal = "Phys. Rev. Lett.",
    volume = "134",
    number = "1",
    pages = "011805",
    year = "2025"
}

@article{XLZD:2024nsu,
    author = "Aalbers, J. and others",
    collaboration = "XLZD",
    title = "{The XLZD Design Book: towards the next-generation liquid xenon observatory for dark matter and neutrino physics}",
    eprint = "2410.17137",
    archivePrefix = "arXiv",
    primaryClass = "hep-ex",
    doi = "10.1140/epjc/s10052-025-14810-w",
    journal = "Eur. Phys. J. C",
    volume = "85",
    number = "10",
    pages = "1192",
    year = "2025"
}

@article{Blum:2016nrz,
    author = "Blum, Kfir and Sato, Ryosuke and Slatyer, Tracy R.",
    title = "{Self-consistent Calculation of the Sommerfeld Enhancement}",
    eprint = "1603.01383",
    archivePrefix = "arXiv",
    primaryClass = "hep-ph",
    doi = "10.1088/1475-7516/2016/06/021",
    journal = "JCAP",
    volume = "06",
    number = "06",
    pages = "021",
    year = "2016"
}

@article{Chang:2010en,
    author = "Chang, Spencer and Weiner, Neal and Yavin, Itay",
    title = "{Magnetic Inelastic Dark Matter}",
    eprint = "1007.4200",
    archivePrefix = "arXiv",
    primaryClass = "hep-ph",
    doi = "10.1103/PhysRevD.82.125011",
    journal = "Phys. Rev. D",
    volume = "82",
    pages = "125011",
    year = "2010"
}

@article{Fitzpatrick:2012ix,
    author = "Fitzpatrick, A. Liam and Haxton, Wick and Katz, Emanuel and Lubbers, Nicholas and Xu, Yiming",
    title = "{The Effective Field Theory of Dark Matter Direct Detection}",
    eprint = "1203.3542",
    archivePrefix = "arXiv",
    primaryClass = "hep-ph",
    doi = "10.1088/1475-7516/2013/02/004",
    journal = "JCAP",
    volume = "02",
    pages = "004",
    year = "2013"
}

@article{Parikh:2024mwa,
    author = "Parikh, Aditya and Sato, Ryosuke and Slatyer, Tracy R.",
    title = "{Regulating Sommerfeld resonances for multi-state systems and higher partial waves}",
    eprint = "2410.18168",
    archivePrefix = "arXiv",
    primaryClass = "hep-ph",
    reportNumber = "MIT-CTP/5790, OU-HET-1243",
    doi = "10.1007/JHEP12(2025)025",
    journal = "JHEP",
    volume = "12",
    pages = "025",
    year = "2025"
}

@book{Raffelt:1996wa,
    author = "Raffelt, G. G.",
    title = "{Stars as laboratories for fundamental physics}: {The astrophysics of neutrinos, axions, and other weakly interacting particles}",
    isbn = "978-0-226-70272-8",
    month = "5",
    year = "1996"
}

@article{BaBar:2014zli,
    author = "Lees, J. P. and others",
    collaboration = "BaBar",
    title = "{Search for a Dark Photon in $e^+e^-$ Collisions at BaBar}",
    eprint = "1406.2980",
    archivePrefix = "arXiv",
    primaryClass = "hep-ex",
    reportNumber = "BABAR-PUB-14-002, SLAC-PUB-15979",
    doi = "10.1103/PhysRevLett.113.201801",
    journal = "Phys. Rev. Lett.",
    volume = "113",
    number = "20",
    pages = "201801",
    year = "2014"
}

@article{Banerjee:2019pds,
    author = "Banerjee, D. and others",
    title = "{Dark matter search in missing energy events with NA64}",
    eprint = "1906.00176",
    archivePrefix = "arXiv",
    primaryClass = "hep-ex",
    reportNumber = "CERN-EP-2019-116",
    doi = "10.1103/PhysRevLett.123.121801",
    journal = "Phys. Rev. Lett.",
    volume = "123",
    number = "12",
    pages = "121801",
    year = "2019"
}

@misc{LDMX:2018cma,
    author = "{\r{A}}kesson, Torsten and others",
    collaboration = "LDMX",
    title = "{Light Dark Matter eXperiment (LDMX)}",
    eprint = "1808.05219",
    archivePrefix = "arXiv",
    primaryClass = "hep-ex",
    reportNumber = "FERMILAB-PUB-18-324-A, SLAC-PUB-17303",
    month = "8",
    year = "2018"
}

@article{Garcia-Cely:2013nin,
    author = "Garcia-Cely, Camilo and Ibarra, Alejandro and Molinaro, Emiliano",
    title = "{Dark matter production  from Goldstone boson interactions and implications for direct searches and  dark radiation}",
    eprint = "1310.6256",
    archivePrefix = "arXiv",
    primaryClass = "hep-ph",
    doi = "10.1088/1475-7516/2013/11/061",
    journal = "JCAP",
    volume = "11",
    pages = "061",
    year = "2013"
}

@article{Berlin:2025fwx,
    author = "Berlin, Asher and Foster, Joshua W. and Hooper, Dan and Krnjaic, Gordan",
    title = "{dSphobic Dark Matter}",
    eprint = "2504.12372",
    archivePrefix = "arXiv",
    primaryClass = "hep-ph",
    reportNumber = "FERMILAB-PUB-25-0227-T",
    month = "4",
    year = "2025",
    journal=""
}

@article{DelaTorreLuque:2025zjt,
    author = "De la Torre Luque, Pedro and Carenza, Pierluca and Nguyen, Thong T. Q.",
    title = "{Sub-keV dark matter can strongly ionize molecular clouds}",
    eprint = "2507.01962",
    archivePrefix = "arXiv",
    primaryClass = "hep-ph",
    month = "7",
    year = "2025",
    journal=""
}

@article{Agius:2025nfz,
    author = "Agius, Dominic and Slatyer, Tracy Robyn",
    title = "{Boosting the cosmic 21-cm signal with exotic Lyman-$α$ from dark matter}",
    eprint = "2510.26791",
    archivePrefix = "arXiv",
    primaryClass = "astro-ph.CO",
    journal="",
    reportNumber = "MIT-CTP/5952",
    month = "10",
    year = "2025"
}

@article{ONeil:2022szc,
    author = "O'Neil, Stephanie and others",
    title = "{Endothermic self-interacting dark matter in Milky Way-like dark matter haloes}",
    eprint = "2210.16328",
    archivePrefix = "arXiv",
    primaryClass = "astro-ph.GA",
    reportNumber = "MIT-CTP/5486",
    doi = "10.1093/mnras/stad1850",
    journal = "Mon. Not. Roy. Astron. Soc.",
    volume = "524",
    number = "1",
    pages = "288--306",
    year = "2023"
}

@article{Bautista:2025lxk,
    author = "Bautista, Yilber Fabian and Robertson, Andrew and Sagunski, Laura and Smith-Orlik, Adam and Tulin, Sean",
    title = "{Jeans Model for the Shapes of Self-interacting Dark Matter Halos}",
    eprint = "2511.10765",
    archivePrefix = "arXiv",
    primaryClass = "astro-ph.CO",
    journal="",
    month = "11",
    year = "2025"
}

\end{document}